\documentclass[numbers,sort,preprint,12pt]{elsarticle}

\usepackage{graphicx}
\usepackage{color}
\usepackage{amsmath}
\usepackage{amssymb}
\usepackage{natmove}
\usepackage{natbib}
\usepackage{hyperref}
\usepackage{bm}
\usepackage{comment}
\usepackage{xurl}
\usepackage[normalem]{ulem}

\newcommand{\D}{\mathcal{D}}
\newcommand{\Y}{\bar{Y}}

\newcommand{\R}{\bm{\mathcal{R}}}
\newcommand{\G}{\mathcal{G}}

\journal{Computer Physics Communications}

\begin{document}
\begin{frontmatter}

\title{A Wigner Matrix Based Convolution Algorithm For Matrix Elements in the LCAO Method} 

\author[aff1,aff2]{Tyler C. Sterling}
\ead{ty.sterling@colorado.edu}
\affiliation[aff1]{
organization={Department of Physics, University of Colorado Boulder},
city={Boulder},
state={CO},
postcode={80309}, 
country={USA}}
\affiliation[aff2]{
organization={Sterling Independent Study and Research},
city={Killeen},
state={TX},
postcode={76549}, 
country={USA}}

\begin{abstract}
The linear combination of atomic orbitals (LCAO) method uses a small basis set in exchange for expensive matrix element calculations. The most efficient approximation for the matrix element calculations is the two-center approximation (2CA) in tight binding (TB). In the 2CA, a variety of matrix elements are neglected with only "two-center integrals" (2CI) remaining. The 2CI are calculated efficiently by rotating to symmetrical coordinates where the integral is parameterized. This makes TB fast in exchange for diminished transferability. An ideal electronic structure method has both the efficiency of TB and the transferability of ab-initio methods. In this work, I expand the full crystal potential into multipoles where the resulting matrix elements are transformed into the form of 2CI between high angular momentum functions. The usual Slater-Koster formulae for TB are limited to $l\leq3$; to enable efficient evaluation of the full crystal potential 2CI, I derive a Wigner matrix based convolution algorithm (WMCA) that works for arbitrary angular momentum. Given a suitable method for generating a local ab-initio Kohn-Sham potential, the algorithm for calculating matrix elements is applicable to fully ab-initio LCAO methods (this is the subject of forthcoming work). In this paper, I apply the WMCA to silicon using a model crystal potential.
\end{abstract}

\begin{graphicalabstract}
\centering
\includegraphics[width=1\linewidth]{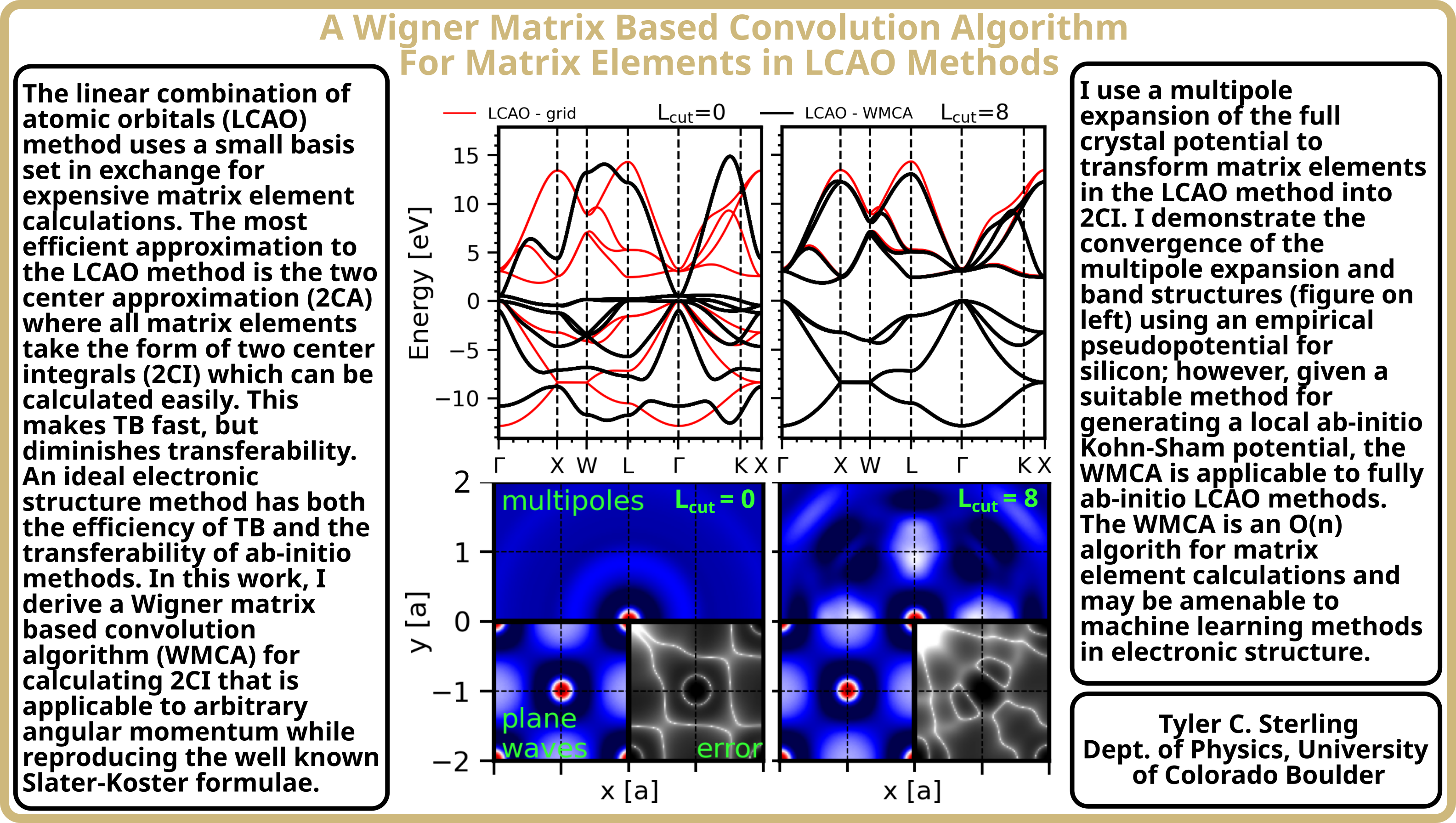}
\end{graphicalabstract}

\begin{highlights}
\item Tight binding like two-center integrals (2CI) are fast to calculate but are restricted to low angular momentum
\item A multipole expansion of the full crystal potential leads to 2CI with high angular momentum
\item A general Wigner matrix based algorithm (WMCA) for 2CI with arbitrary angular momentum is derived
\item The WMCA is applicable, in principle, to ab-initio all electron LCAO calculations
\end{highlights}


\end{frontmatter}

\section{Introduction}

The linear combination of atomic orbitals (LCAO) method uses a small (i.e. efficient) basis set at the cost of expensive matrix element calculations. One of the most efficient approximations for the matrix element calculations is the two-center approximation (2CA) in tight binding (TB) \cite{slater1954simplified}. In the 2CA, the crystal potential is approximated as a sum of spherical atom-centered functions and only LCAO matrix elements involving two distinct sites are kept; all matrix elements between pairs of orbitals reduce to "two-center integrals" (2CI) which can be parameterized and calculated efficiently. The tradeoffs for the speed of the 2CA are i) the skill and effort required to parameterize the Hamiltonian and ii) decreased transferability. A total energy/molecular dynamics method with the efficiency of TB in the 2CA that is as transferable as ab-initio methods and requires only minimal parameters is ideal.

The canonical method \cite{slater1954simplified,lendi1974extension,takegahara1980slater} for evaluating each 2CI is to rotate the orbital quantization axes to point along the internuclear separation vector; in the rotated coordinates, there is azimuthal symmetry and each 2CI is easy to evaluate (or fit). This is very efficient and, as such, even advanced forms of TB that provide total energies and forces (e.g. DFTB \cite{elstner1998self,elstner2014density,spiegelman2020density,hourahine2020dftb+,hourahine2010dftb+}, xTB \cite{bannwarth2019gfn2,bannwarth2021extended,grimme2017robust}) ultimately use the 2CA for the band structure part of the total energy. The 2CI parameters and the rest of the total energy are approximated with relatively simple analytical forms that are efficient to calculate, but require careful fitting and/or calculation of many parameters.

If we can transform the matrix elements of the \textit{full} crystal potential into the form of 2CI, we keep all contributions to the potential matrix elements (two-center, three-center, etc.), minimizing the need for parameterization while retaining the efficiency of the 2CI formalism. To that end, I expand the crystal potential in multipoles around each atom in the unit cell; the product of an orbital and multipole centered on the same site factorizes into a sum over functions with well defined angular momentum. The integral over each term in the sum is a separate 2CI that can be treated efficiently, in principle, by rotating to azimuthally symmetric coordinates. In the symmetric coordinates, the remaining integral is a convolution that can be calculated using Fourier transforms and the convolution theorem. This is similar in spirit to Becke's multi-center expansion \cite{becke1988multicenter} and related methods \cite{delley1990all,hirshfeld1977bonded,delley1996fast,blum2009ab} for atom-centered integration, with the key difference being the use of multipoles. The use of multipoles enables application of the well established machinery for angular momentum in quantum mechanics, circumventing numerical angular integrations in matrix element calculations. Notably, this algorithm is applicable, in principle, to fully ab-initio all-electron calculations.

Historically, TB methods \cite{slater1954simplified,lendi1974extension,takegahara1980slater} rotate orbitals using precalculated tables of formulae for the rotation coefficients. This is fast from a computational point of view, but the explicit derivation of the tables of rotation coefficients is increasingly tedious and error prone for increasing angular momentum, with tables for $f$ electrons \cite{takegahara1980slater,lendi1974extension,sharma1979general,podolskiy2004compact,durgavich2016extension} appearing decades after the original tables up to $d$ orbitals were published by Slater and Koster \cite{slater1954simplified}. There are only a few codes available that implement the canonical method for $f$ electrons \cite{durgavich2016extension,hourahine2020dftb+,hourahine2010dftb+,bursch2017fast} and likely no codes exist that implement it for higher angular momentum. Even if the expansion is truncated at $l=2$ in the matrix elements of the full crystal potential, for a system with $d$ electrons we need rotation coefficients up to $l=4$ which, to the best of my knowledge, don't exist in tables anywhere. An alternative, programmatic algorithm for calculating the rotation coefficients that is applicable to arbitrary angular momentum is needed. 

Programmatic approaches to 2CI that depend on Wigner rotation matrices have been proposed before \cite{sharma1979general,podolskiy2004compact} but haven't caught on because, ultimately, unfavorable explicit formulas are used for the Wigner little-$d$ matrix; this problem is well known in e.g nuclear physics \cite{wang2022effective,tajima2015analytical}, quantum metrology \cite{pezze2007phase,pezze2008mach}, and elsewhere \cite{miyazaki2007wigner,yang2012franck,li2025enhancing,zhang2022equivariant}. Only recently has an algorithm for calculating Wigner matrices that is stable for all $l$ been published \cite{feng2015high}. It is based on the diagonalization of the Pauli-$y$ matrix and requires no factorial evaluations. This new "universal" algorithm enables general application of Wigner matrices to 2CI with arbitrary angular momentum. 

In what follows, I present a simple and efficient Wigner matrix based convolution algorithm (WMCA) for calculating 2CI between arbitrary angular momentum functions. The Wigner matrices are calculated using a universal algorithm \cite{feng2015high} that is applicable to arbitrary angular momentum. By analytical calculation of the Wigner matrices for small angular momentum, I show that the tabulated SK formulae \cite{slater1954simplified} are reproduced if the 2CA is assumed, validating the WMCA. I then expand the full crystal potential into multipoles centered on each atom in the unit cell. The product of each multipole and orbital is a sum over orbital-like functions with well defined angular momentum and the resulting matrix element are 2CI, some with high angular momentum. All of the 2CI are evaluated using the WMCA. Given a suitable algorithm for generating a local ab-initio Kohn-Sham potential, the WMCA algorithm for calculating matrix elements is applicable to any ab-initio LCAO method, where it is potentially an $O(n)$ algorithm for synthesizing the matrix elements. Moreover, the local atomic representation of the potential in the WMCA may be amenable to machine learning methods in electronic structure calculations. Application to ab-initio LCAO calculations is the subject of forthcoming work. In this paper, I apply the WMCA to silicon using a model crystal potential. 

\section{LCAO Formalism}

As a starting point, I introduce the "linear combination of atomic orbitals" (LCAO) method. I use a general Hamiltonian that is a sum of the kinetic energy operator and a crystal potential; in an ab-initio calculation, this could be e.g. the Kohn-Sham Hamiltonian. I solve it by choosing LCAO functions as a basis for calculating the Hamiltonian and overlap matrix elements \cite{slater1954simplified,kohanoff2006electronic} and then diagonalizing the secular equation \cite{kohanoff2006electronic,thijssen2007computational} to determine the eigenvalues. The LCAO basis functions are defined as
\begin{equation}\begin{gathered}
    \varphi_{\vec k \mu}(\vec r) = c_{\vec k \mu} \sum_{\vec R} e^{i\vec k \cdot (\vec R + \vec \tau_\mu)} \phi_\mu(\vec r - (\vec R + \vec \tau_\mu)) 
\end{gathered}\end{equation} 
with $\vec k$ is a $k$-point in the first Brillouin zone, $\vec R$ labels the origin of the unit cell, and $\vec \tau_i$ labels the position of the $i^{\rm th}$ atom in the unit cell. $\phi_\mu(\vec r )$ is a pseudoatomic orbital consisting of a real spherical harmonic, $X_{lm}(\Omega_{\vec r})$, times a localized radial function, $\chi_{\mu}( r)$: $\phi_\mu(\vec r) = X_{lm}(\Omega_{\vec r}) \chi_\mu(r)$. Since $X_{lm}(\Omega_{\vec r})$ is real, I pick $\phi_\mu(\vec r)$ to be real by choosing $\chi_{\mu}(\vec r)$ to be real functions. $c_{\vec k \mu}$ is a normalization coefficient defined such that $\langle \varphi_{\vec k \mu} | \varphi_{\vec k \mu} \rangle = 1$. The radial functions $\chi_{\mu}(r)$ are different for each orbital type but only depend on angular momentum number $l$ and shell $n$ for each type. $\mu$ is a composite index labeling atom $i$, shell $n$, and angular momentum $l,m$: $\mu=\{i,n,l,m\}$. The LCAO functions are an appropriate basis because they satisfy Bloch's theorem as can be shown easily.

An overlap matrix element is defined as 
\begin{equation}\begin{gathered}
    S_{\mu\nu}(\vec k) = \langle \varphi_{\vec k \mu}|\varphi_{\vec k \nu}\rangle =  \bar c_{\vec k \mu} c_{\vec k \nu}   e^{i\vec k \cdot (\vec \tau_\nu  - \vec \tau_\mu)} \sum_{\vec R} e^{i\vec k \cdot \vec R} S_{\mu\nu}(\vec \Delta) \\
    S_{\mu\nu}(\vec \Delta) \equiv \int d\vec r  \phi_\mu(\vec r) \phi_\nu(\vec r -  \vec \Delta)  
    \label{eq:overlap_matrix_element}
\end{gathered}\end{equation}
where $\vec \Delta = \vec R +(\vec \tau_\nu-\vec \tau_\mu)$ has orbital indices $\mu,\nu$ implied by the matrix element subscripts. The over-bar on $\bar c_{\vec k \mu}$ means complex conjugation. A Hamiltonian matrix element is 
\begin{equation}\begin{gathered}
    H_{\mu\nu}(\vec k) = \langle \varphi_{\vec k \mu}| \bm H |\varphi_{\vec k \nu}\rangle =  \bar c_{\vec k \mu} c_{\vec k \nu}  e^{i\vec k \cdot (\vec \tau_\nu  - \vec \tau_\mu)} \sum_{\vec R} e^{i\vec k \cdot \vec R} H_{\mu\nu}(\vec \Delta) \\
    H_{\mu\nu}(\vec \Delta) \equiv \int d\vec r  \phi_\mu(\vec r) \bm H \phi_\nu(\vec r - \vec \Delta) .  
    \label{eq:hamiltonian_matrix_element}
\end{gathered}\end{equation}
The kinetic energy and overlap matrix elements are "two-center integrals" (2CI), i.e. integrals over two localized functions centered on different sites. As a function of internuclear separation, they are convolutions. They are relatively easy to calculate using Wigner matrices and the convolution theorem; below, I provide a general method for calculating these convolutions that works for arbitrary angular momentum. 

The most difficult part of the Hamiltonian matrix element is the potential:
\begin{equation}\begin{gathered}
    V_{\mu\nu}(\vec \Delta) = \int d\vec r  \phi_\mu(\vec r) V(\vec r) \phi_\nu(\vec r - \vec \Delta) .  
    \label{eq:potential_matrix_element}
\end{gathered}\end{equation}
In general, the crystal potential can't be represented in a convenient analytical form. Calculating matrix elements in ab-initio full potential methods typically requires representing the potential on a dense grid or doing atom centered angular integrations numerically \cite{delley1990all,becke1988multicenter,hirshfeld1977bonded,delley1996fast,blum2009ab}. The matrix elements of the full potential are the most computationally demanding part of an LCAO calculation. 

\subsection{The two-center Approximation}

I now introduce the ubiquitous two-center approximation (2CA) \cite{slater1954simplified}. This both motivates the algorithm for calculating 2CI presented below and serves as a benchmark to validate against. The 2CA first replaces the crystal potential by an empirical pseudopotential that is a sum of spherical atom-centered functions \cite{slater1954simplified}:
\begin{equation}
\begin{gathered}
    V(\vec{r}) = \sum_{\vec{R}i} v_i(|\vec{r}-(\vec{R}+\vec{\tau}_i)|) .
    \label{eq:spherical_potential}
\end{gathered}
\end{equation}
The subscript $i$ labels the $i^{\rm{th}}$ atom in the unit cell and $\vec R$ is the coordinate vector of the origin of the unit cell. $v_i(|\vec{r}-(\vec{R}+\vec{\tau}_i)|)$ is a spherical pseudopotential centered on the atom located at $\vec{R}+\vec{\tau}_i$. A potential matrix element is
\begin{equation}
\begin{gathered}
    V_{\mu\nu}(\vec \Delta) = \sum_{\vec{R}' i'} \int d\vec{r} \phi_{\mu}(\vec{r}) v_{i'}(\vec{r}-(\vec{R}'+\vec{\tau}_{i'})) \phi_{\nu} (\vec{r}-\vec \Delta) .
\end{gathered}
\end{equation}
This isn't a drastic approximation yet since, if the empirical potential is an accurate representation of the full crystal potential, the matrix elements are accurate. However, the 2CA is drastic. Call the sites on which the bra orbital, the potential, and the ket orbital are centered $\vec{A}$, $\vec{B}$, and $\vec{C}$ respectively. The integral is $\int d\vec{r} \phi_{\mu} (\vec{r}-\vec{A}) V(\vec{r}-\vec{B}) \phi_\nu (\vec{r}-\vec{C})$. In the 2CA \cite{slater1954simplified}, we neglect three-center terms with $\vec{A} \neq \vec{B} \neq \vec{C}$ and crystal field terms with $\vec A = \vec C \neq \vec B$, keeping only on site ($\vec A = \vec B = \vec C$) and two-center ($\vec A = \vec B \neq \vec C$ and $\vec C = \vec B \neq \vec A$) terms. The justification is that there are a huge number of three-center and crystal field terms and they are hard to calculate; if we assume that the effects of the three-center terms can be described by \emph{effective} two-center matrix elements, then we drastically reduce the work we have to do. 

In the 2CA, each potential matrix element is a 2CI which is the same form as the kinetic energy and overlap matrix elements; these are very fast to calculate. The 2CI are evaluated by rotating to azimuthally symmetric coordinates where the integral is simpler and is usually fit or parameterized. The caveat is the canonical method for rotating is restricted to $l\leq 3$ \cite{takegahara1980slater,lendi1974extension,sharma1979general}. I now derive a generalized algorithm for calculating 2CI that works for arbitrary angular momentum via Wigner matrices and enables application of the 2CI algorithm to matrix elements of the full crystal potential.

\section{Wigner Matrix Convolution Algorithm for two-center Integrals}\label{sec:convolution_algorithm}

\begin{figure}
    \centering
    \includegraphics[width=0.5\linewidth]{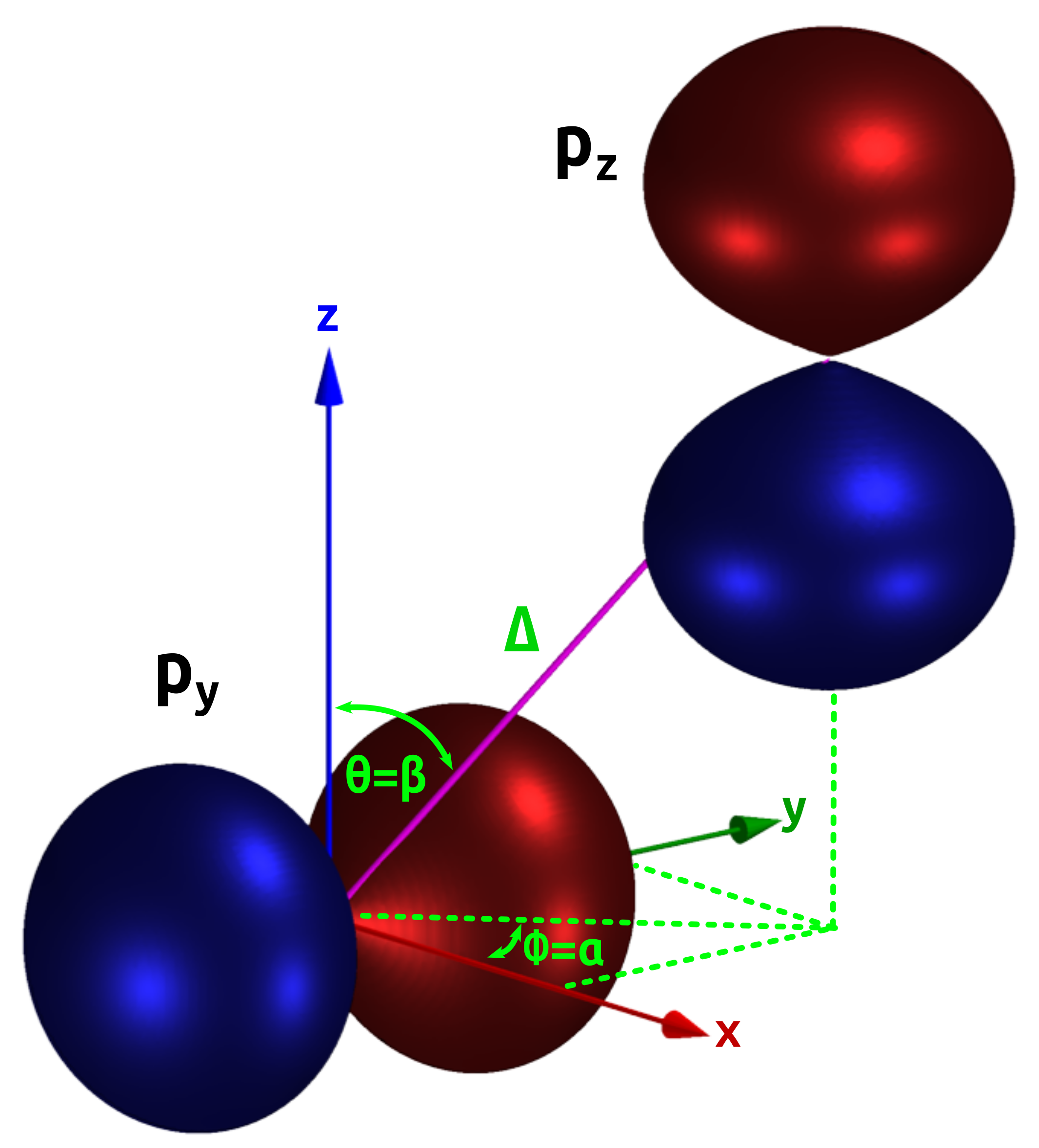}
    \caption{A $p_y$ and $p_z$ orbital are centered at the origin and the vector $\vec{\Delta}$. The azimuthal and polar angles are $\phi$ and $\theta$ and the internuclear distance is $\Delta = |\vec{\Delta}|$. Since there is no "roll" angle associated with a coordinate vector, the Euler angles for a space-fixed $ZYZ$ rotation are (i) $\alpha=\phi$ around the $z$ axis and (ii) $\beta=\theta$ around the $y$ axis.}
    \label{fig:fig_1}
\end{figure}

To demonstrate how to calculate the 2CI, I solve a slightly more general problem of the convolution between two arbitrary localized functions in three dimensions. Call them $a(\vec r)$ and $b(\vec r)$. "Localized" means decaying to zero at large distances. I calculate the convolutions using a Wigner matrix based convolution algorithm (WMCA). I expand the convolution into integrals over pairs of functions with well defined angular momentum by developing a multipole expansion of each function as 
\begin{equation}\begin{gathered}
    a(\vec r) = \sum_{lm} X_{lm}(\Omega_{\vec r})a_{lm}(r) \\
    a_{lm}(r) = \int d\Omega_{\vec r} X_{lm}(\Omega_{\vec r}) a (\vec r)
\end{gathered}\end{equation}
and similarly for $b(\vec r)$. I use real spherical harmonics since these are a natural basis for pseudoatomic orbitals (they are the $s$, $p$, $d$, $f$, and so on atomic orbital functions). $\Omega_{\vec r}$ is the solid angle associated with unit vector $\hat r = \vec r/r$.

The convolution of $a(\vec r)$ and $b(\vec r)$ is defined as
\begin{equation}\begin{gathered}
    C(\vec \Delta) = \int d\vec r a(\vec r) b(\vec r - \vec \Delta)
\end{gathered}\end{equation} 
where the integral is over all space and $\vec \Delta$ is an arbitrary position vector. Since the functions are localized, the boundary conditions are irrelevant and $C(\vec \Delta)$ is localized and only needs to be calculated for a finite domain, making this problem tractable. Using the multipole expansion:
\begin{equation}\begin{gathered}
    C(\vec \Delta) = \sum_{lm}\sum_{l'm'} C_{lml'm'} (\vec \Delta)
\end{gathered}\end{equation}
which, for well behaved functions, is a relatively quickly convergent sum over angular momenta. Each term,
\begin{equation}\begin{gathered}
    C_{lml'm'}(\vec \Delta) =  \int d\vec r  X_{lm}(\Omega_{\vec r}) a_{lm}( r) X_{l'm'}(\Omega_{\vec r - \vec \Delta})  b_{l'm'}(|\vec r - \vec \Delta |),
    \label{eq:two_center_integral}
\end{gathered}\end{equation}
is a convolution between two localized functions with well defined angular momentum (cf. Fig. \ref{fig:fig_1}). Eq. \ref{eq:two_center_integral} is a ``two-center integral" (2CI). The kinetic energy and overlap matrix elements and, after expanding the crystal potential in multipoles, the matrix elements of each angular momentum channel of the potential are 2CI in the form of eq. \ref{eq:two_center_integral}. 

\begin{figure*}
    \centering
    \includegraphics[width=1\linewidth]{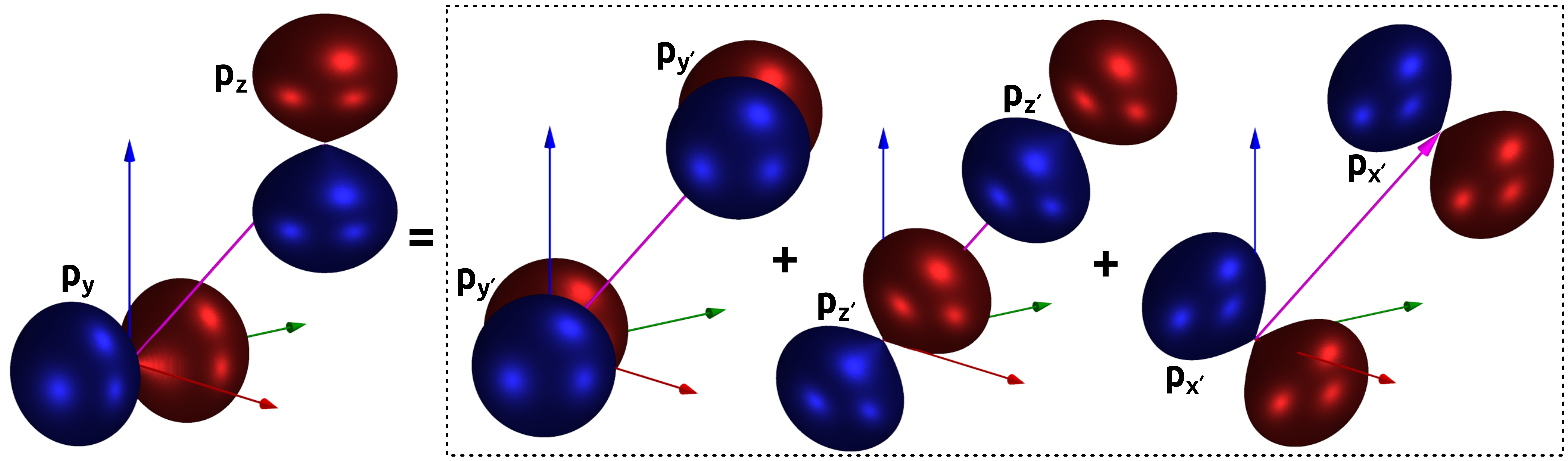}
    \caption{The overlap integral between a $p_y$ and $p_z$ orbital with internuclear vector $\vec{\Delta}$ (cf. Fig. \ref{fig:fig_1}) can be decomposed into three azimuthally symmetric integrals by rotating the coordinate system so that $\vec{\Delta}=\Delta \vec{z}'$. The azimuthal symmetry conserves magnetic quantum number $M$ in the new coordinate system so that only three integrals survive, with $M=M' = -1 = p_{y'}$, $M=M' = 0 = p_{z'}$, and $M=M' = 1 = p_{x'} $. $p_{x'}$ and $p_{y'}$ are perpendicular to the vector and $p_{z'}$ is parallel to it. In the conventional notation, $M$ is denoted by $\sigma = 0$,  $\pi = \pm1$, and $\delta = \pm 2$.}
    \label{fig:fig_2}
\end{figure*}

We can simplify the integral in eq. \ref{eq:two_center_integral} by rotating the spherical harmonics so that $\vec{\Delta} \rightarrow \Delta \vec{z}$ points along the $\vec{z}$ axis in the rotated basis and then exploiting azimuthal symmetry (cf. Fig. \ref{fig:fig_2}). The real spherical harmonics transform under rotation as
\begin{equation}\begin{gathered}
    X_{lm}(\Omega_{\vec r-\vec \Delta}) = \sum_{m'} \D^{l}_{mm'}(\hat \Delta) X_{lm'}(\Omega_{\vec r - \Delta \vec z})
    \label{eq:real_sph_rotation}
\end{gathered}\end{equation} 
with $\Omega_{\vec r-\vec \Delta}$ and $\Omega_{\vec r- \Delta\vec z}$ the angles along the original and new axes respectively specified by vector $\vec \Delta$. $\D^{l}_{mm'}(\hat \Delta)$ is the Wigner-$\D$ matrix element for real spherical harmonics with rotation operator $\R_{\hat \Delta}$. The matrices $\bm \D^{l}(\hat \Delta)$ are real by definition. The real Wigner-$\D$ matrices are related to the usual complex Wigner-$D$ matrices; the details, and the algorithm for calculating the Wigner-$\D$ matrices, are presented in \ref{appendix:wigner_matrices}. The convolutions become
\begin{equation}\begin{gathered}
    C_{lml'm'}(\vec \Delta) =\sum_{M} \D^l_{mM}(\hat \Delta) \D^{l'}_{m'M }(\hat \Delta) \Gamma_{M}^{lml'm'}(\Delta) \\
    \Gamma_{M}^{lml'm'}(\Delta) = \int d\vec r a_{lm}(r) X_{lM}(\Omega_{\vec r}) b_{l'm'}(|\vec r - \Delta \vec z |)X_{l'M}(\Omega_{\vec r- \Delta \vec z}) 
    \label{eq:clm}
\end{gathered}\end{equation}
where the integral is in coordinates where the angular parts of $a_{lm}(\vec r)$ and $b_{l'm'}(\vec r-\vec \Delta)$ share a common azimuthal variable. The azimuthal symmetry constrains the magnetic quantum numbers on the spherical harmonics to be the same and $|M|\leq\min(l,l')$. To see this, consider the explicit form of the real spherical harmonics \cite{blanco1997evaluation}: $X_{lm}(\Omega)=\Theta_{l|m|}(\theta)\Phi_{m}(\phi)$. The polar function $\Theta_{lm}(\theta) = N_{lm}P_l^m(\cos\theta)$ is the same as in the complex spherical harmonics; $P_l^m(\cos \theta)$ are associated Legendre polynomials and $N_{lm}$ are the normalization coefficients. The difference is in the azimuthal part, $\Phi_m(\phi)$:
\begin{equation}\begin{gathered}
\Phi_{m}(\phi) = 
\begin{cases}
    \sqrt{2} \cos( m \phi) \quad & \textrm{if} \quad m>0 \\
    1 \quad & \textrm{if} \quad m = 0 \\
    \sqrt{2} \sin(|m|\phi) \quad & \textrm{if} \quad m<0 .
\end{cases}
\end{gathered}\end{equation} 
In the rotated frame, the integral is proportional to $\int_0^{2\pi} d\phi \Phi_{M}(\phi)\Phi_{M'}(\phi) = 2\pi \delta_{MM'}$. This constrains the azimuthal quantum number to be the same in the rotated coordinate system (see Fig. \ref{fig:fig_2}). When writing eq. \ref{eq:clm} above, I have factored out the Kronecker delta, $\delta_{MM'}$, but not the $2\pi$ since the integral over $\phi$ is still included.

I now use the convolution theorem to calculate the $\Gamma$-integrals. Define $a_{lm}^M(\vec r) = a_{lm}(r) X_{lM}(\Omega_{\vec r}) $ and similarly for $b_{l'm'}^{M}(\vec r - \Delta \vec z )$. Their Fourier transforms are 
\begin{equation}\begin{gathered}
    a_{lm}^M(\vec r) = \int \frac{d\vec q}{(2\pi)^{3/2}} a_{lm}^M(\vec q) e^{i\vec q \cdot \vec r} \quad \quad a_{lm}^M(\vec q) = \int \frac{d\vec r}{(2\pi)^{3/2}} a_{lm}^M(\vec r) e^{-i\vec q \cdot \vec r}\\
    a_{lm}^M(\vec q) =  \sqrt{\frac{2}{\pi}} \sum_{L'M'} (-i)^{L'} X_{L'M'}(\Omega_{\vec q}) \int d\vec r a_{lm}^M(\vec r) j_{L'}(qr) X_{L'M'}(\Omega_{\vec r}) 
\end{gathered}\end{equation}
and similarly for $b_{l'm'}(\vec r - \Delta \vec z )$. I used the plane wave expansion identity in the supp. info \cite{supp}. Then 
\begin{equation}\begin{gathered}
    a_{lm}^M(\vec q) =  (-i)^{l} X_{lM}(\Omega_{\vec q}) A_{lm}(q) \\
    A_{lm}(q) \equiv \sqrt{\frac{2}{\pi}} \int_0^\infty dr r^2 j_{l}(qr) a_{lm}(r) 
\end{gathered}\end{equation}
and similarly for $b_{l'm'}^M(\vec q)$. $j_{l}(qr)$ is a spherical Bessel function of degree $l$ and so on. $A_{lm}(q)$ and $B_{l'm'}(q)$ are spherical Bessel transforms. Since the radial functions are localized, the spherical Bessel transforms are convergent.

The $\Gamma$-integral becomes
\begin{equation}\begin{gathered}
    \Gamma^{lml'm'}_{M}(\Delta) =  \int d\vec r a_{lm}^M(\vec r) b_{l'm'}^M(\vec r - \Delta \vec z) = \int d\vec q a_{lm}^M(-\vec q) b_{l'm'}^M(\vec q) e^{-i\Delta \vec q \cdot \vec z}  \\
    = \sqrt{2} \pi \sum_{L} i^{l-l'-L} \sqrt{2L+1} \G_{lMl'M}^{L0} \Lambda^{lml'm'}_{L}(\Delta) 
    \label{eq:gamma}
\end{gathered}\end{equation}
where 
\begin{equation}\begin{gathered}
    \Lambda^{lml'm'}_{L}(\Delta) \equiv \sqrt{\frac{2}{\pi}} \int_0^\infty dq q^2 j_{L}(q\Delta) A_{lm}(q) B_{l'm'}(q) 
    \label{eq:lambda}
\end{gathered}\end{equation}
is the inverse spherical Bessel transform of the product $A_{lm}(q)$ and $B_{l'm'}(q)$; in other words, $\Lambda^{lml'm'}_{L}(\Delta)$ is an application of the convolution theorem using spherical Bessel transforms. $\G_{lMl'M}^{L0}$ is a Gaunt coefficient for real spherical harmonics (the Gaunt coefficients can be calculated easily, see the supp. info. \cite{supp}). I used the fact that $X_{LM}(\Omega_{-\vec q}) = (-1)^{L} X_{LM}(\Omega_{\vec q}) $ and $X_{LM}(\Omega_{\vec z}) = \delta_{M0} \sqrt{(2L+1 )/ 4\pi}$. The sum over $L$ is restricted to $|l-l'| \leq L \leq l+l'$ and $l+l'+L$ an even integer by the selection rules of the real Gaunt coefficients. $\Gamma^{lml'm'}_{M}(\Delta)$ and $\Lambda^{lml'm'}_{L}(\Delta)$ are always real since $l-l'-L$ is either 0 ($i^{0}=1$) or an even integer ($i^{2n}=-1$). The bulk of the work in calculating eq. \ref{eq:gamma} is calculating the forward and inverse Bessel transforms.

\section{Two-center Integrals: Matrix Elements in the Two-center Approximation}

\begin{figure}
    \centering
    \includegraphics[width=0.5\linewidth]{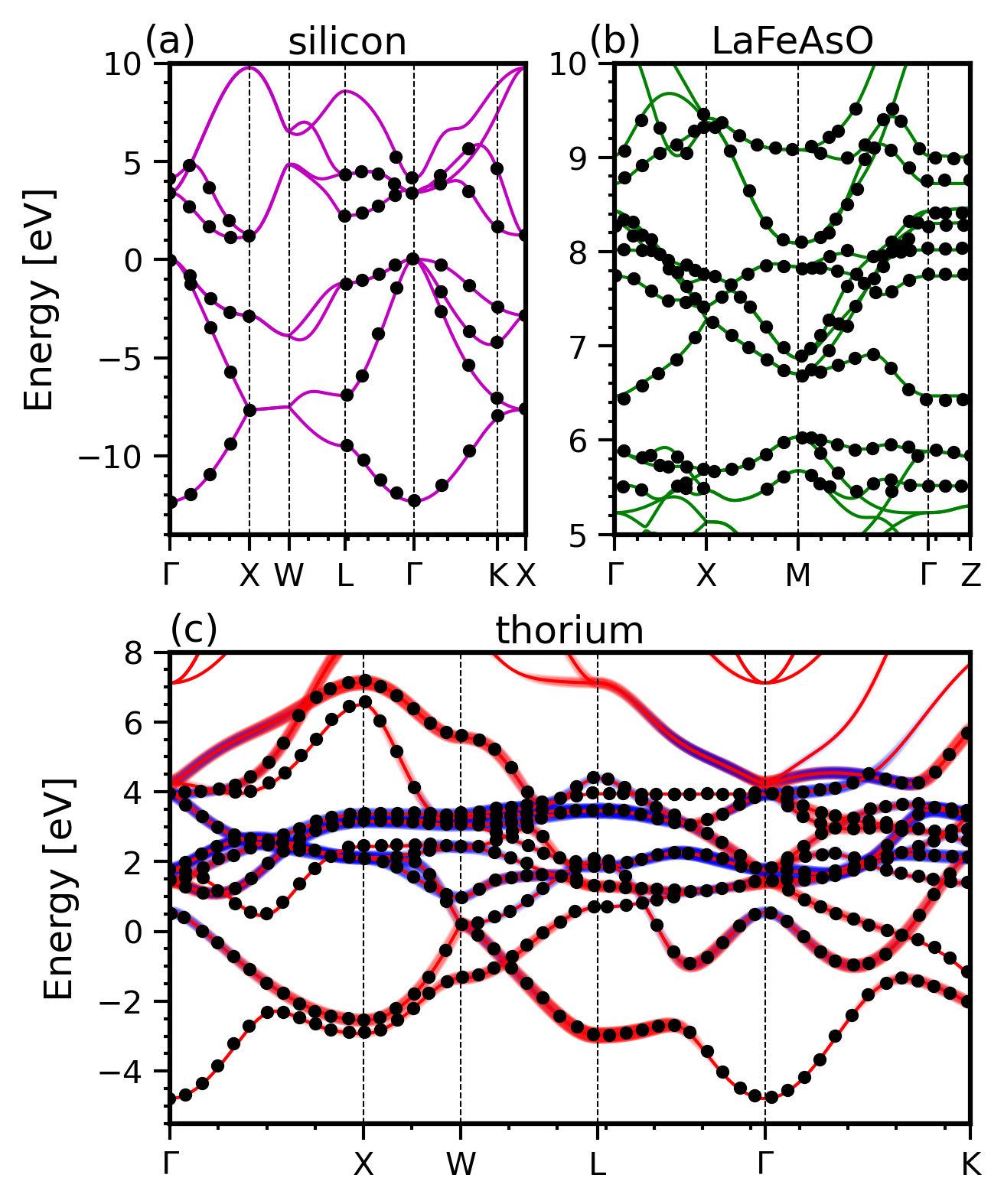}
    \caption{Band structures of (a) silicon in the fcc-diamond phase \cite{mattheiss1981electronic} and (b) tetragonal LaFeAsO \cite{papaconstantopoulos2010tight} calculated in SK 2CA and (c) of fcc thorium using the NRL method \cite{durgavich2016extension,mehl1996applications,cohen1994tight}. The parameters were taken from the references and the band structures were calculated using the WMCA. The silicon data points in (a) are the pseudopotential results \cite{chelikowsky1974electronic} that the band structure was fit to. In (b) and (c), the data points are the digitized tight binding band structure calculated with the same parameters in the refs. \cite{papaconstantopoulos2010tight,durgavich2016extension}. My calculation in (c) is projected onto the $d$ (red) and $f$ (blue) orbitals, showing the $f$ character of the electrons near the Fermi level ($E=0$ eV) in thorium.}
    \label{fig:band_structure}
\end{figure}

I now validate the WMCA for 2CI by calculating matrix elements in the 2CA; both the Hamiltonian and overlap matrix elements are in the form of eq. \ref{eq:two_center_integral}. Approximations for the $\Gamma$-integral can be derived from the Kohn-Sham energy functional with parameters calculated from ab-initio methods \cite{elstner1998self,elstner2014density,grimme2017robust,bannwarth2019gfn2,bannwarth2021extended} or the $\Gamma$-integrals can be treated as fitting parameters like in the SK scheme. In the SK scheme, they are the well-known $(ss\sigma)$, $(sp\pi)$, $(pd\pi)$, and so-on parameters with $\sigma$ labeling orbitals with $M=0$, $\pi$ with $M= \pm1$, and $\delta$ with $M=\pm 2$. A SK Hamiltonian matrix element is
\begin{equation}\begin{gathered}
    H_{\mu \nu}(\vec \Delta) = \sum_M \D^l_{mM}(\hat \Delta)\D^{l'}_{m'M}(\hat \Delta) (ll'M)
    \label{eq:sk_2CI}
\end{gathered}\end{equation}
which is identical to an overlap matrix element with different parameters. As I now show, the product of Wigner-$\D$ matrix elements generates the "direction cosines" terms in the Slater-Koster tables. The algorithm for calculating Wigner-$\D$ matrices, similar to Reference \cite{feng2015high}, is in \ref{appendix:wigner_matrices}. A similar algorithm, specifically for parameterized SK 2CI, has been proposed before in the appendix of Reference \cite{martin2020electronic}.

I want to calculate 2CI between two $s$, an $s$ and $p$, and two $p$ orbitals and compare to the known formulas in table 1 in Reference \cite{slater1954simplified}. The $\bm \D^{0}(\hat \Delta) = 1$ matrix is trivial, so I only need to calculate the $\bm \D^{1}(\hat \Delta)$ matrix. The analytical form of the $\bm \D^{1}(\hat \Delta)$ matrix is derived in \ref{appendix:analytical}:
\begin{equation}
\begin{gathered}
    \bm \D^{1}(\alpha,\beta,0) = 
    \begin{pmatrix} c_\alpha & s_\alpha s_\beta &  s_\alpha c_\beta  \\
    0 & c_\beta & - s_\beta  \\
    - s_\alpha  &  c_\alpha  s_\beta &  c_\alpha c_\beta 
    \end{pmatrix} 
\end{gathered}
\end{equation}
where I use shorthand $\cos\alpha=c_\alpha$, $\sin\alpha=s_\alpha$, and so on. For rotations in TB, we are rotating a vector and there is no need for a $\gamma$ rotation associated with a rigid body roll. Then the azimuthal and polar angles, $\phi=\alpha$ and $\theta=\beta$, equal the Euler angles (see Fig. \ref{fig:fig_1}). 

These matrix elements are related to the direction cosines by $a= \vec{\Delta}\cdot \vec{x}/\Delta = c_\alpha s_\beta$, $b = \vec{\Delta}\cdot \vec{y}/\Delta = s_\alpha s_\beta$, and $c = \vec{\Delta}\cdot \vec{z}/\Delta = c_\beta$. Obviously, $S_{s,s}(\vec{\Delta}) = (ss\sigma)$. For $s$ and $p$ orbitals, $S_{s,px}(\vec{\Delta}) = a (sp\sigma) \equiv E_{s,x}$. For $p$ orbitals (cf. Fig. \ref{fig:fig_2}),
\begin{equation}
\begin{gathered}
    S_{px,px}(\vec{\Delta})
    = a^2 (pp\sigma) + (1-a^2) (pp\pi) \equiv E_{x,x} \\
    S_{px,py}(\vec{\Delta})  
    = ab(pp\sigma) - ab (pp\pi )\equiv E_{x,y} \\
    S_{px,pz}(\vec{\Delta}) 
    = ac (pp\sigma) - ac (pp\pi) \equiv E_{x,z} .
\end{gathered}
\end{equation}
These, and $E_{s,x}$, agree with Table 1 in Reference \cite{slater1954simplified}.

I implemented the WMCA in \ref{appendix:wigner_matrices} in a code to calculate 2CI in the 2CA using eq. \ref{eq:two_center_integral} and tested it on several materials: fcc-diamond silicon including $s$ and $p$ orbitals \cite{mattheiss1981electronic} is a classical example (Fig. \ref{fig:band_structure}a), tetragonal LaFeAsO including Fe $d$ orbitals and As and O $p$ orbitals \cite{papaconstantopoulos2010tight} is a complex, multi-species system with diverse coordination, and thorium \cite{durgavich2016extension} includes $f$ orbitals. For all materials, I used parameters from the literature and the WMCA to calculate the rotation coefficients. The agreement is perfect.

\section{Two-center Integrals: Kinetic Energy and Overlap Matrix Elements}\label{sec:ke_ovlp}

I now wish to go beyond the 2CA and apply the WMCA in Section \ref{sec:convolution_algorithm} to matrix elements assuming arbitrary LCAO basis functions and a full crystal potential (i.e. no 2CA or shape approximation). To that end, the kinetic energy 2CI are 
\begin{equation}\begin{gathered}
    T_{\mu\nu}(\vec \Delta) = \int d\vec r  \phi_\mu(\vec r) \left( -\frac{1}{2} \nabla^2 \right) \phi_\nu(\vec r - \vec \Delta) = \sum_{M} \D^l_{mM}(\hat \Delta) \D^{l'}_{m'M}(\hat \Delta) t^{\mu\nu}_M(\Delta) \\
    t^{\mu\nu}_M(\Delta) = \int d\vec r  \phi^M_\mu(\vec r)  \left( -\frac{1}{2} \nabla^2 \right) \phi^M_\nu(\vec r -  \Delta \vec z)  .
\end{gathered}\end{equation}
with $\phi_\mu^M(\vec r) \equiv \chi_\mu(r) X_{lM}(\Omega_{\vec r})$ and so on. Recall that $\mu=\{i,n,l,m\}$ and $\nu=\{j,n',l',m'\}$ are composite indices with $l$ the angular momentum of orbital $\phi_\mu(\vec r)$ and $l'$ the angular momentum of orbital $\phi_\nu(\vec r)$. The $t^{\mu\nu}_M(\Delta)$ integral is a convolution which, after Fourier transforming, becomes
\begin{equation}\begin{gathered}
    t^{\mu\nu}_M( \Delta) = \frac{1}{2} \int d\vec q q^2 \phi^M_\mu(-\vec q) \phi^M_\nu(\vec q) e^{-i\Delta \vec q \cdot \vec z}  .
\end{gathered}\end{equation}
Then
\begin{equation}\begin{gathered}
    t^{\mu\nu}_M(\Delta) = \sqrt{2} \pi  \sum_{L} i^{l-l'-L} \sqrt{2L+1} \G_{lMl'M}^{L0} \alpha_L^{\mu\nu}(\Delta) \\
    \alpha_L^{\mu\nu}(\Delta) \equiv  \sqrt{\frac{2}{\pi}}\int_0^\infty dq q^2   j_L(q\Delta) \left[\frac{q^2}{2}  \xi^{\mu}_l(q)  \xi^{\nu}_{l'}(q) \right].
\end{gathered}\end{equation}
Overlap 2CI are similarly 
\begin{equation}\begin{gathered}
    S_{\mu\nu}(\vec \Delta) = \int d\vec r  \phi_\mu(\vec r)  \phi_\nu(\vec r - \vec \Delta) = \sum_{M} \D^l_{mM}(\hat \Delta) \D^{l'}_{m'M}(\hat \Delta) s^{\mu\nu}_M(\Delta) \\
    s^{\mu\nu}_M(\Delta) = \int d\vec r  \phi^M_\mu(\vec r) \phi^M_\nu(\vec r -  \Delta \vec z)  .
\end{gathered}\end{equation}
Then
\begin{equation}\begin{gathered}
    s^{\mu\nu}_M(\Delta) = \sqrt{2}\pi \sum_{L} i^{l-l'-L} \sqrt{2L+1} \G_{lMl'M}^{L0} \beta_L^{\mu\nu}(\Delta) \\
    \beta_L^{\mu\nu}(\Delta) \equiv \sqrt{\frac{2}{\pi}} \int_0^\infty dq q^2   j_L(q\Delta) \xi^{\mu}_l(q)  \xi^{\nu}_{l'}(q) 
\end{gathered}\end{equation}
with 
\begin{equation}\begin{gathered}
    \xi^{\mu}_{l}(q) \equiv \sqrt{\frac{2}{\pi}} \int_0^\infty dr r^2 j_{l}(qr) \chi_\mu(r) ,
    \label{eq:orbital_bessel_transform}
\end{gathered}\end{equation}
the spherical Bessel transform of the isolated orbital radial function. Since $l-l'-L$ must be an even integer, $t_M^{\mu\nu}(\Delta)$, $\alpha_L^{\mu\nu}(\Delta)$, $s_M^{\mu\nu}(\Delta)$, and $\beta_L^{\mu\nu}(\Delta)$ are real. The inverse spherical Bessel transforms, $\alpha_L^{\mu\nu}(\Delta)$ and $\beta_L^{\mu\nu}(\Delta)$, can be calculated once and for all and at the beginning of a calculation and interpolated to arbitrary distance as needed. The only difference between kinetic energy and overlap matrix elements is a factor of $2$ and the power of the wave number in the inverse Bessel transform.

\section{Two-center Integrals: Matrix Elements of the Full Crystal Potential}

Now I want to transform the full crystal potential into the form of 2CI so that I can apply the WMCA in Section \ref{sec:convolution_algorithm}. To that end, I expand the full crystal potential in multipoles centered on each site. This is key to this method for two reasons: i) similar in spirit to the ubiquitous multi-center expansion of Becke \cite{delley1990all,becke1988multicenter,hirshfeld1977bonded,delley1996fast,blum2009ab}), we circumvent calculating three-center integrals and, ii) by representing the potential with functions of well defined angular momentum, it enables the use of Wigner matrices to simplify integrals. 

\subsection{Multipole Expansion of the Potential}

In each potential matrix element, eq. \ref{eq:potential_matrix_element}, the full crystal potential is defined to be centered on $\phi_\mu(\vec r)$, i.e. the origin. The multipole expansion of the crystal potential has a different representation in each matrix element, but each representation is the same potential so this is valid. In terms of real spherical harmonics, the multipole expansion around atom $i$ (where $\phi_\mu(\vec r)$ resides) is 
\begin{equation}\begin{gathered}
    V(\vec r) = \sum_{LM}^{L_\textrm{cut}} V^i_{LM}(r) X_{LM}(\Omega_{\vec r}) \\
    V^i_{LM}(r) = \int_\Omega d\Omega_{\vec r} X_{LM}(\Omega_{\vec r}) V(\vec r)
    \label{eq:ace_potential}
\end{gathered}\end{equation}
where $L_\textrm{cut}$ is a finite angular momentum cutoff. Since each orbital on atom $i$ sees the same crystal potential, the multipole expansion only needs to be calculate around each \textit{atom}, not around each \textit{orbital}. The superscript $i$ is used to label the potential expansion around the atoms; each orbital on atom $i$ sees the same crystal potential multipoles represented by $V^i_{LM}(r)$.

I assume the crystal potential is a local, periodic function that extends over all space. Strictly speaking, since the extent of the potential is infinite, there is no upper bound on the angular momentum. However, consider how coordination increases with increasing distance; at small distances, coordination is low and the potential barely changes when rotating by a small amount. At large distance, the coordination is huge and a small change in angle results in a rapid change in the potential. As a result, the crystal potential converges relatively quickly with increasing $L$ at short distances, with only large $L$ components at large distances (see below). Since the orbitals are localized, the product of two orbital radial functions is small (or vanishing for truly "local" orbitals \cite{soler2002siesta,lewis2011advances,garcia2020siesta,blum2009ab}) unless the orbitals are close together. This establishes a "cutoff" distance beyond which matrix elements can be neglected; then the potential only needs to be calculated accurately within the cutoff distance, putting an \emph{effective} upper bound on the angular momentum and enabling us to use a finite $L_\textrm{cut}$.

What remains to develop the multipole expansion are the details of evaluating the angular integrals in eq. \ref{eq:ace_potential}. Two such algorithms are presented in this paper: below, I present a formula for calculating the multipoles from an arbitrary potential on a grid. This is applicable, in principle, to fully ab-initio dual-space LCAO calculations that use a plane wave representation of the potential and density \cite{vandevondele2005quickstep,lippert1997hybrid,kuhne2020cp2k}. In the supplementary information \cite{supp}, I present a real space algorithm that is applicable to an empirical crystal potential that is a sum of spherical atom centered functions (but doesn't use the 2CA or other shape approximation).

\subsection{Multipole Expansion from a Local Potential on a Grid}\label{sec:plane_wave_multipole}

Assume that we have stored a self-consistently converged local potential on a grid: this could be from e.g. from an all-electron LDA or GGA self-consistent calculation or from an empirical pseudopotential. Calculating the multipoles from a potential represented a grid is straightforward in reciprocal space. The Fourier expansion of the potential is defined as
\begin{equation}\begin{gathered}
    V(\vec r) = \sum_{\vec G} V_{\vec G} e^{i\vec G \cdot \vec r} \\
    V_{\vec G} = \frac{1}{\mathcal V} \int_{\mathcal V} d\vec r V(\vec r) e^{-i\vec G \cdot \vec r} 
\end{gathered}\end{equation}
where the integral is over a single unit cell with $\mathcal V$ the unit cell volume and $\vec G$ are reciprocal lattice vectors. Using eq. \ref{eq:ace_potential} and the plane wave expansion formula for real spherical harmonics (see the supp. info. \cite{supp}),
\begin{equation}\begin{gathered}
    V^i_{LM}(r) = \sum_{\vec G} V_{\vec G} \int d\Omega_{\vec r} X_{LM}(\Omega_{\vec r}) e^{i\vec G \cdot \vec r} = 4\pi i^L \sum_{\vec G} V_{\vec G} j_L(Gr) X_{LM}(\Omega_{\vec G})
    \label{eq:ace_reciprocal_space}
\end{gathered}\end{equation}
Given the potential on a reciprocal space grid, we can immediately calculate $V^i_{LM}(r)$ from eq. \ref{eq:ace_reciprocal_space}; given the potential in real space, we can first fast Fourier transform to reciprocal space and then use eq. \ref{eq:ace_reciprocal_space}. 

While the sum over reciprocal lattice vectors, $\vec G$, in eq. \ref{eq:ace_reciprocal_space} is straightforward to evaluate, the number of grid points (i.e. number of $\vec G$) scales with system size, so for large unit cells the sum becomes expensive. In the spirit of the WMCA for 2CI, it is desirable to have a real-space algorithm that calculates the full potential out to a cutoff without using plane waves (i.e. independent of system size). This is the subject of ongoing work. 

\subsection{Potential Matrix Elements}\label{sec:potential_matrix_elements}

A potential matrix element between two orbitals and the multipole representation of the potential is 
\begin{equation}\begin{gathered}
    V_{\mu\nu}(\vec \Delta) = \sum_{LM}^{L_\textrm{cut}} V^{LM}_{\mu\nu}(\vec \Delta)  
\end{gathered}\end{equation}
with
\begin{equation}\begin{gathered}
    V^{LM}_{\mu\nu}(\vec \Delta) \equiv \int d\vec r  \phi_\mu(\vec r) V^i_{LM}(r) X_{LM}(\Omega_{\vec r})  \ \phi_\nu(\vec r - \vec \Delta) = \\
    \int d\vec r \chi_\mu(r)  V^i_{LM}(r) X_{lm}(\Omega_{\vec r}) X_{LM}(\Omega_{\vec r})  \chi_\nu(|\vec r - \vec \Delta|)  X_{l'm'}(\Omega_{\vec r - \vec \Delta}).  
\end{gathered}\end{equation}
Each angular momentum channel of the potential contributes separately to the matrix element, so we treat each $V_{\mu\nu}^{LM}(\vec \Delta)$ term separately. We can use the product formula for real spherical harmonics (see the supp. info. \cite{supp}), $X_{lm}(\Omega_{\vec r})X_{LM}(\Omega_{\vec r}) = \sum_{L'M'} \G_{lmLM}^{L'M'} X_{L'M'}(\Omega_{\vec r})$, and rewrite
\begin{equation}\begin{gathered}
    V^{LM}_{\mu\nu}(\vec \Delta)  = \sum_{L'M'} \G_{lmLM}^{L'M'} 
    \int d\vec r \chi_\mu(r)  V^i_{LM}(r) X_{L'M'}(\Omega_{\vec r})  \chi_\nu(|\vec r - \vec \Delta|)  X_{l'm'}(\Omega_{\vec r - \vec \Delta}) .
    \label{eq:full_potential_matrix_element}
\end{gathered}\end{equation}
with $\G_{lmLM}^{L'M'}$ a Gaunt coefficient in the basis of real spherical harmonics. The integral is now in the form of a 2CI like eq. \ref{eq:two_center_integral} and we can use the machinery in sec. \ref{sec:convolution_algorithm} to evaluate each one efficiently. Since we truncate $L \leq L_\textrm{cut}$ and $L'$ is always close to $L$ (i.e. $|l-L| \leq L' \leq l+L$, with $l \lesssim3$), each matrix element of the separate angular momentum channels is a relatively small sum over 2CI. The only caveat is that the angular momentum $L'$ is often much larger than $L'=3$, the largest angular momentum that can be practically calculated using the SK lookup tables \cite{takegahara1980slater,lendi1974extension,sharma1979general}. Thus, a universal algorithm for calculating Wigner matrices with arbitrary angular momentum, like that presented in \ref{appendix:wigner_matrices}, is essential. Using the Wigner matrices
\begin{equation}\begin{gathered}
    V^{LM}_{\mu\nu}(\vec \Delta) = \sum_{L'M'} \G_{lmLM}^{L'M'} \sum_{M''} 
    \D^{L'}_{M'M''}(\hat \Delta) \D^{l'}_{m'M''}(\hat \Delta) \Gamma_{L'M''}^{\mu\nu LM}(\Delta) \\
    \Gamma_{L'M''}^{\mu\nu LM}(\Delta) \equiv \int d\vec r \chi_\mu(r)  V^i_{LM}(r) X_{L'M''}(\Omega_{\vec r})  \chi_\nu(|\vec r - \Delta \vec z|)  X_{l'M''}(\Omega_{\vec r - \Delta \vec z}) .
    \label{eq:V_LM}
\end{gathered}\end{equation}
I need $\Gamma_{L'M''}^{\mu\nu LM}(\Delta)$ for all internuclear distances $\Delta$. In practice, $\Gamma_{L'M''}^{\mu\nu LM}(\Delta)$ is calculated on a grid and interpolated to arbitrary $\Delta$. I calculate it with the convolution theorem: 
\begin{equation}\begin{gathered}
    \Gamma_{L'M''}^{\mu\nu LM}(\Delta) =  \int d\vec r \Psi^{\mu LM}_{L'M''}(\vec r) \psi^{\nu}_{M''}(\vec r - \Delta \vec z) = \int d\vec q \Psi^{\mu LM}_{L'M''}(-\vec q) \psi^{\nu}_{M''}(\vec q) e^{-i\Delta \vec q \cdot \vec z}  \\
    = \sqrt{2} \pi \sum_{l''} i^{L'-l'-l''} \sqrt{2l''+1} \G_{L'M''l'M''}^{l''0} \Lambda_{l''L'}^{\mu\nu LM}(\Delta) 
    \label{eq:gamma_pot}
\end{gathered}\end{equation}
with
\begin{equation}\begin{gathered}
    \Lambda_{l''L'}^{\mu\nu LM}(\Delta) \equiv \sqrt{\frac{2}{\pi}} \int_0^\infty dq q^2 j_{l''}(q\Delta) \Xi^{\mu LM}_{L'}(q) \xi^\nu_{l'}(q) 
    \label{eq:lambda_pot} 
\end{gathered}\end{equation}
the inverse spherical Bessel transform of the product of the spherical Bessel transforms of the potential-weighted orbital radial function,
\begin{equation}\begin{gathered}
    \Xi^{\mu LM}_{L'}(q) \equiv \sqrt{\frac{2}{\pi}} \int_0^\infty dr r^2 j_{L'}(qr) \chi_\mu(r)  V^i_{LM}(r) ,
    \label{eq:big_xi} 
\end{gathered}\end{equation}
and the spherical Bessel transform of the isolated orbital radial function, $\xi_{l'}^\nu(q)$, defined in eq. \ref{eq:orbital_bessel_transform}.

Given orbitals and a potential, the bulk of the work is calculating the Forward and inverse Bessel transforms for each $L'$ and $l''$. Fortunately the angular momentum numbers $L'$ and $l''$ are restricted to a relatively small set (of relatively large angular momentum) by the selection rules. In some cases, the $\xi^\nu_{l'}(q)$ transform can be calculated analytically: e.g. for spherical Gaussian orbitals, a closed form for the integral in eq. \ref{eq:orbital_bessel_transform} is known \cite{kuang1997molecular}. The $\Xi^{\mu LM}_{L'}(q)$ integral can't be calculated analytically in general since $V^i_{LM}(r)$ are numerical functions on a radial grid. Since we are doomed to calculate the forward Bessel transforms numerically, there is no reason to restrict ourselves to otherwise analytically convenient basis functions like Slater and Gaussian functions and we are free to use more computationally efficient numerical orbitals \cite{soler2002siesta,lewis2011advances,garcia2020siesta,blum2009ab}.

\section{Results}

\begin{figure}
    \centering
    \includegraphics[width=0.5\linewidth]{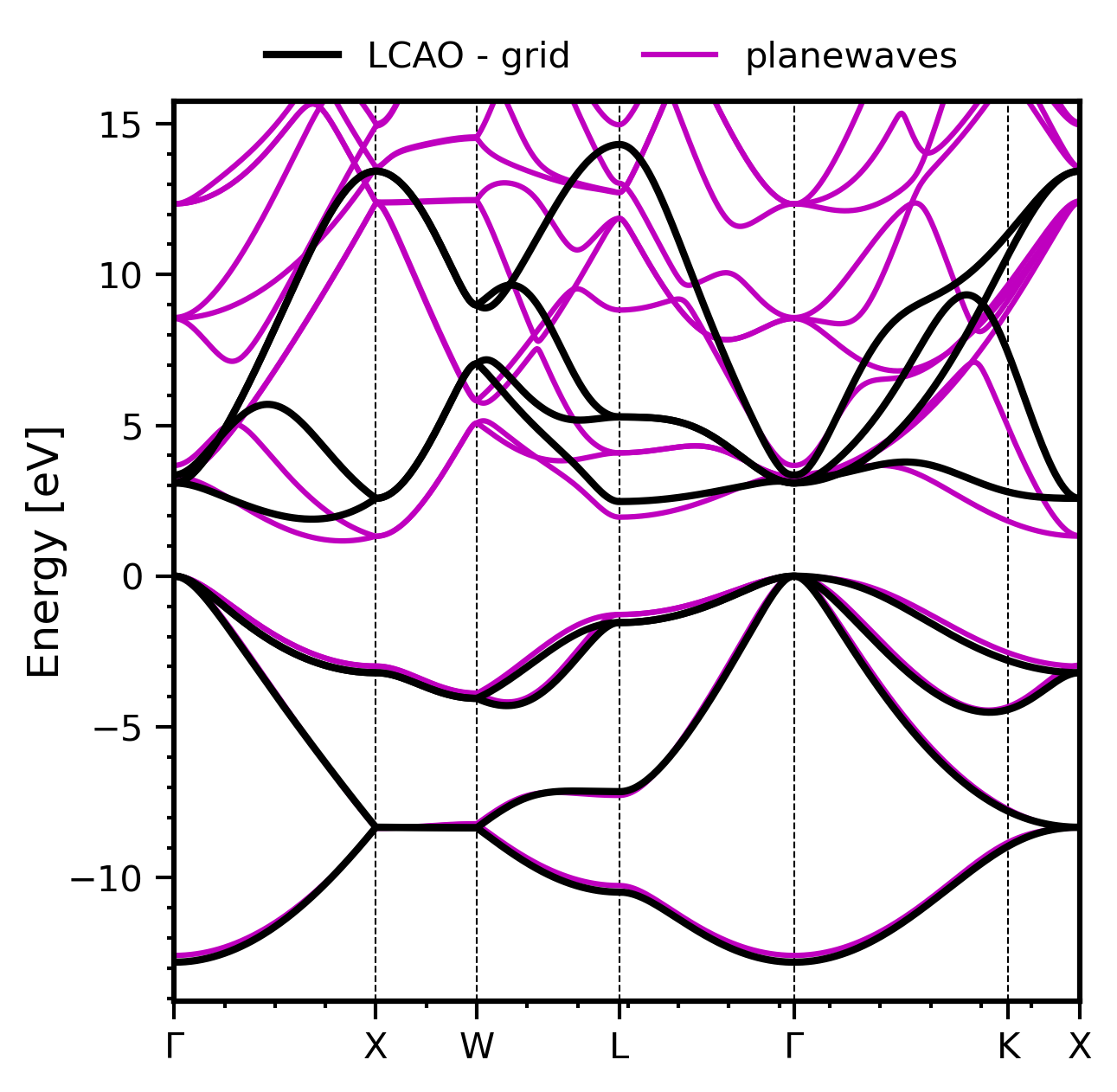}
    \caption{Comparison between the band structure of silicon calculated using the minimal silicon SZV-MOLOPT-SR-GTH basis \cite{vandevondele2007gaussian} on a grid (black lines) and using a plane wave basis (magenta lines); both calculations used the empirical pseudopotential of Wang \textit{et al.} \cite{wang1994electronic}.}
    \label{fig:pw_vs_lcao}
\end{figure}

The most important parameter in determining the accuracy of the multipole expansion of the potential, and the corresponding potential matrix elements, is the angular momentum cutoff, $L_\textrm{cut}$. In the matrix element calculation, the computational cost of increasing other parameters, e.g. the number of radial grid points, is relatively low and I assume from here on that the results are converged with respect to all other quantities. Here, I want to check two main things with respect to $L_\textrm{cut}$: i) I want to validate the convergence of the multipole expansion of the potential in real space and ii) I want to validate the accuracy of WMCA by calculating band structures. While the general methods presented here should work for any well behaved numerical basis functions and crystal potential, I pick simple examples from the literature. 

I study silicon: for the pseudoatomic orbital basis functions, I use spherical (as opposed to Cartesian) contracted Gaussian basis functions \cite{kuang1997molecular,kohanoff2006electronic,vandevondele2007gaussian}:
\begin{equation}\begin{gathered}
    \phi_{\mu}(\vec r) = X_{lm}(\Omega_{\vec r}) \chi_\mu(r) \\
    \chi_{\mu}(r) = r^l \sum_j c_{\mu}^j e^{-\alpha^{j}_\mu r^2} 
\end{gathered}\end{equation}
with each $\alpha^j_{\mu}$ and $c_{\mu}^j$ a different exponent and contraction coefficient respectively. The exponents and contraction coefficients are from the silicon SZV-MOLOPT-SR-GTH basis set \cite{vandevondele2007gaussian} of CP2K \cite{kuhne2020cp2k}. This minimal basis set serves as a simple example; in the supp. info. \cite{supp}, I repeat the following calculations using the TZVP-MOLOPT-GTH basis set. For the pseudopotential, I use an empirical, local function \cite{wang1994electronic,chelikowsky1974electronic}: 
\begin{equation}\begin{gathered}
    V(\vec r) = \sum_{\vec R a} v_{Si}(|\vec r - (\vec R + \vec \tau_{a})|) \equiv \sum_{\vec G} V_{\vec G} e^{i\vec G \cdot \vec r} \\
    V_{\vec G} = \frac{v_{Si}(G)}{\mathcal{V}}  \sum_{a} e^{-i\vec G \cdot \vec \tau_{a}}
    \label{eq:local_pseudo}
\end{gathered}\end{equation}
with $\mathcal{V}$ the unitcell volume. I use the parameterization of Wang \textit{et. al} \cite{wang1994electronic}:
\begin{equation}\begin{gathered}
    v_{Si}(G) = \frac{a_1(G^2-a_2)}{a_3e^{a_4G^2}-1} 
\end{gathered}\end{equation}
with $a_1= 36.262$, $a_2=2.19$, $a_3=2.06$, and $a_4=0.487$ in Hartree atomic units. In the supplementary information \cite{supp}, I calculate the multipole expansion of this pseudopotential in real space. I emphasize that, while the pseudopotential in eq. \ref{eq:local_pseudo} assumes that the crystal potential can be represented by spherical functions centered on the atoms, it does not assume the 2CA or other shape approximation. 

I wrote a program that calculates the multipole expansion from the plane wave grid representation of the potential (section \ref{sec:plane_wave_multipole}) and calculates the kinetic energy and overlap matrix elements (section \ref{sec:ke_ovlp}) and the potential matrix elements (section \ref{sec:potential_matrix_elements}) using the WMCA. To validate the results, I also put the LCAO basis functions on a plane wave grid \cite{kuang1997molecular} and calculated the matrix elements by direct numerical integration with the plane wave potential. The details of calculating matrix elements between LCAO basis functions in the plane wave representation are presented elsewhere \cite{louie1979self}. All results from the LCAO grid calculations were converged with respect to plane wave energy cutoff; as such, they are a benchmark with which to compare the matrix elements calculated using the WMCA.

Wang's empirical pseudopotential was designed for a converged planewave basis set; in contrast, I am using a minimal LCAO basis. To ensure that the LCAO calculation is sensible with this potential, I compare the converged LCAO grid calculation to the band structure calculated using a converged plane waves basis set in Figure \ref{fig:pw_vs_lcao}. Ultimately, the valence bands are in good agreement while the conduction bands are in poorer but decent agreement. This is expected with a minimal basis that is optimized to describe occupied states and I conclude that Wang's empirical pseudopotential is suitable to use with silicon SZV-MOLOPT-SR-GTH basis set \cite{vandevondele2007gaussian}.

\begin{figure*}
    \centering
    \includegraphics[width=1\linewidth]{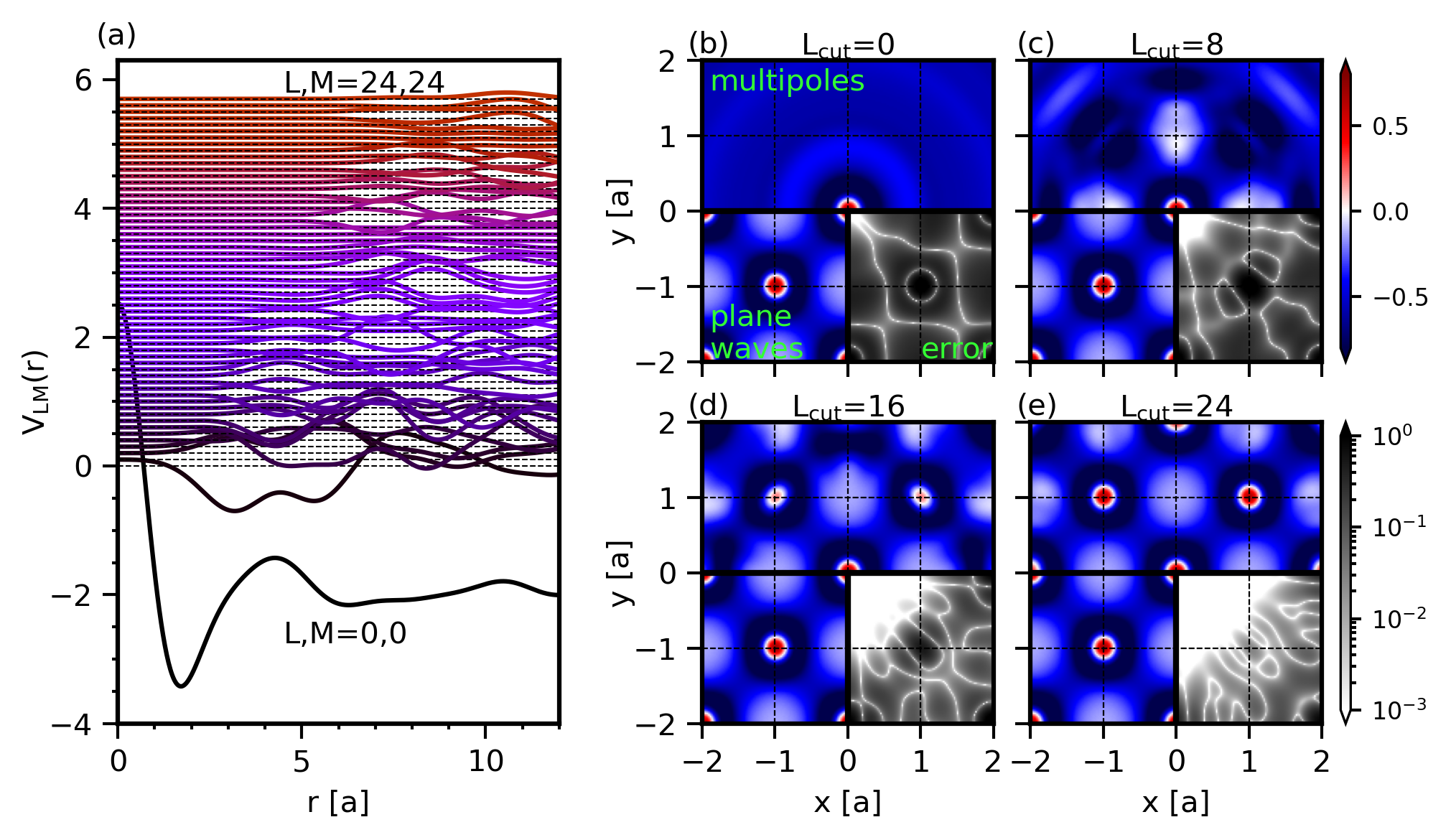}
    \caption{Convergence of the multipole expansion of the silicon crystal potential with increasing angular momentum cutoff, $L_\textrm{cut}$. The conventional 8-atom unit cell is used ($a=5.43~\textrm{\AA}$). Radial functions with amplitude $\max |V_{LM}(r)| \geq 0.1 \times \max|V_{00}(r)|$ up to $(L,M)=(24,24)$ are shown in (a). The radial functions are offset vertically for clarity; the zero of each curve is represented by a dashed line. The truncated potential evaluated on a grid in the $x-y$ plane (top) is compared to the exact potential (lower left) in (b-d). The error, $|V(\vec r) - \sum_{LM} V_{LM}(r)X_{LM}(\Omega_{\vec r})|$, is shown on the lower right in (b-e). Note that the error is shown in log scale. }
    \label{fig:sph_convergence}
\end{figure*}

\subsection{The Multipole Expansion}

The convergence of the multipole expansion of Wang's empirical pseudopotential \cite{wang1994electronic} to the converged plane wave grid potential is checked as a function of $L_\textrm{cut}$ in Figure \ref{fig:sph_convergence}. The multipole expansion of the potential converges relatively quickly with increasing $L_\textrm{cut}$ close to the origin (Fig.  \ref{fig:sph_convergence}a). High angular momentum radial functions ($L\sim 24$) are only large at large distances.

Figures \ref{fig:sph_convergence} (b-e) compare the potential on a real space grid using the multipole expansion (top) to the converged plane wave representation (lower left). The absolute value of the error is smallest at short distances from the center of the multipole expansion and increases at large distances (lower right). For large $L_\textrm{cut}$, the multipole expansion describes the potential quite accurately within a sphere centered on the origin.

\begin{figure}
    \centering
    \includegraphics[width=0.5\linewidth]{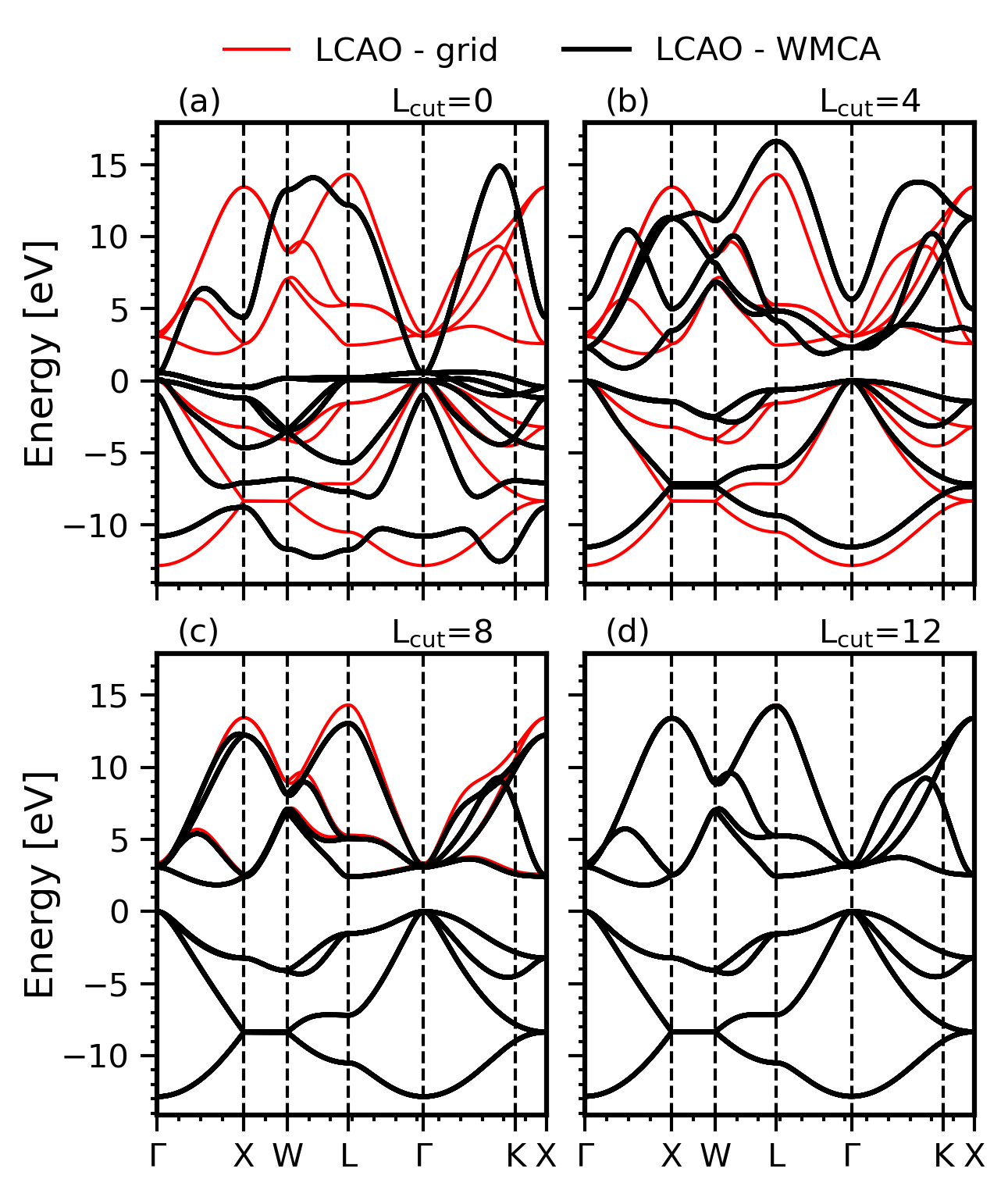}
    \caption{Comparison between the band structures of silicon using the minimal SZV-MOLOPT-SR-GTH basis \cite{vandevondele2007gaussian} and the WMCA (black lines) and using the grid representation of the basis functions (red lines). The band structure using the WMCA is calculated for various potential angular momentum cutoffs, $L_\textrm{cut}$. The Fermi energy is set to zero in all calculations.}
    \label{fig:wigmat_vs_grid}
\end{figure}

\subsection{Band Structures}

\begin{figure}
    \centering
    \includegraphics[width=0.5\linewidth]{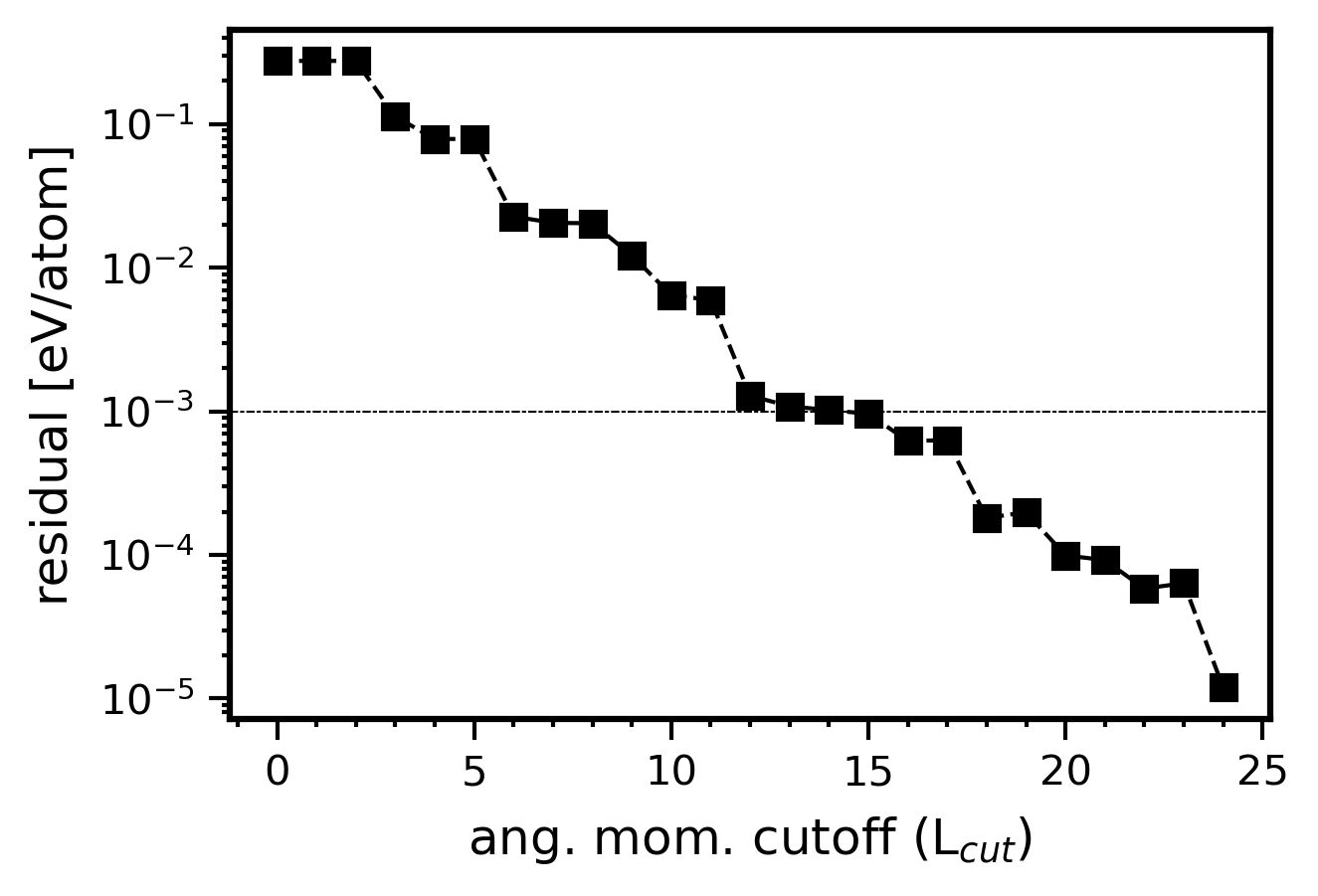}
    \caption{Convergence of the band structure calculated as a function of potential angular momentum cutoff, $L_\textrm{cut}$, using the WMCA to the band structure calculated using a converged grid representation of the LCAO basis functions. The convergence is defined in the text. The horizontal line indicates meV/atom accuracy.}
    \label{fig:wigmat_convergence}
\end{figure}

The convergence of the band structure using the WMCA is validated against to the converged grid calculation in Figures \ref{fig:wigmat_vs_grid} and \ref{fig:wigmat_convergence}. All bands are in relatively poor agreement with $L_\textrm{cut}=0$ and the system is metallic. With $L_\textrm{cut}=4$, the valence bands are approaching good agreement and the system is insulating as it should be. Already for $L_\textrm{cut}=8$, the valence bands are in very good agreement and only the highest energy conduction bands are still in modest disagreement. For $L_\textrm{cut}=12$, the band structure is indistinguishable from the converged LCAO grid calculation. 

To quantify the agreement, I calculate the residual of the band structures defined as 
\begin{equation}\begin{gathered}
    R(L_\textrm{cut}) = \frac{1}{N_{\vec k}N_B N_A} \sqrt{\sum_{\vec k n} \left( \epsilon^\textrm{WMCA}_{\vec k n}(L_\textrm{cut}) - \epsilon^\textrm{grid}_{\vec k n} \right)^2}
\end{gathered}\end{equation}
with $\epsilon^\textrm{WMCA}_{\vec k n}(L_\textrm{cut})$ the WMCA eigenvalues, $\epsilon^\textrm{grid}_{\vec k n}$ the grid calculation eigenvalues, and $N_{\vec k}$, $N_B$, and $N_A$ the number of $\vec k$-points, bands, and atoms respectively. The accuracy converges nearly exponentially with deviations coming from the non-uniform convergence of the potential with angular momentum cutoff (certain $L$ are small or forbidden by symmetry, so the accuracy of the multipole expansion isn't improved by including them).

\begin{figure}
    \centering
    \includegraphics[width=0.5\linewidth]{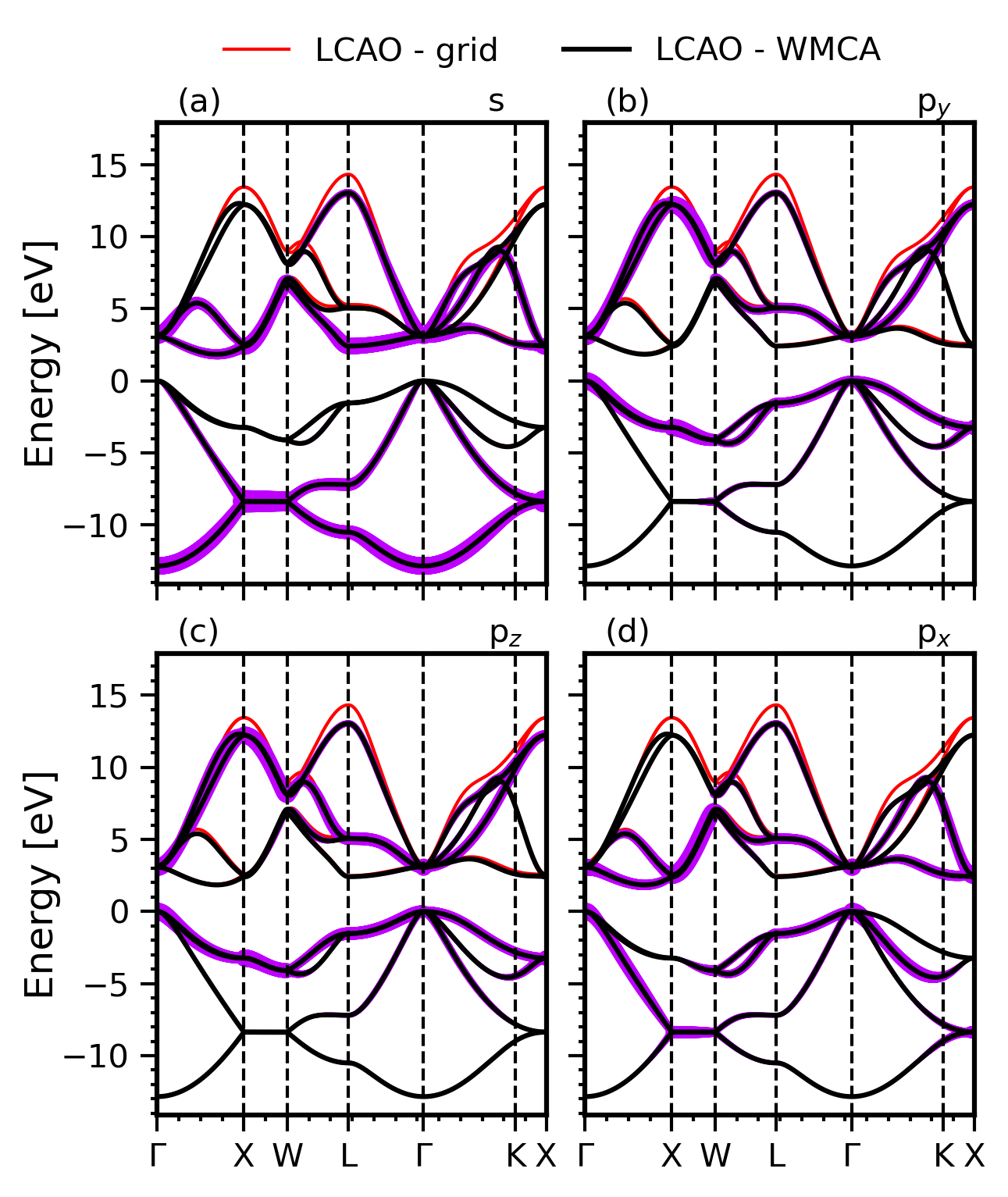}
    \caption{Projections of the pseudoatomic orbitals onto the band structure calculated with the WMCA and $L_\textrm{cut}=8$ (black lines). The width of the purple lines indicates the magnitude of the projections. The LCAO grid calculation is shown in red. Bands composed mostly of the more diffuse $p$ orbitals are in poorer agreement. The Fermi energy is set to zero in all calculations.}
    \label{fig:projections}
\end{figure}

The accuracy of the WMCA band structures with respect $L_\textrm{cut}$ depends strongly on the spatial extent of the pseudoatomic basis functions. At $L_\textrm{cut}=8$, the valence bands are well converged with only high energy conduction bands in error (Fig. \ref{fig:wigmat_vs_grid}c). Projecting the bands with $L_\textrm{cut}=8$ onto pseudoatomic orbitals shows that the bands in worst agreement have predominately $p$ orbital character (Fig. \ref{fig:projections}b-d); the radial part of the $p$ orbitals is more diffuse than that of the $s$ orbital so the matrix elements between $p$ orbitals require the multipole expansion of the potential to be converged to further distance (i.e. larger $L_\textrm{cut}$). 

\section{Discussion}

The fact that the multipole expansion of the potential can be calculated from an arbitrary local potential on a grid is notable because it demonstrates the applicability of the WMCA, at least in principle, to the band structure part of ab-initio, full potential, all electron LCAO density functional theory calculations. This might be useful in existing and future LCAO codes for several reasons: first, the accuracy of matrix elements can be systematically improved to within desired tolerances by increasing $L_\textrm{cut}$. 

Moreover, a hierarchy of numerical approximations can be devised to tune the speed/accuracy of calculations. A particularly attractive example is the ability to use a large $L_\textrm{cut}$, while "Fourier" filtering multipoles with small radial functions. This can substantially reduce the cost of calculating matrix elements, especially if large $L_\textrm{cut}$ components are filtered, without biasing the accuracy of the potential close to the origin. Another example is the ability to tune the cutoff separately for each atom; i.e. we can tune speed vs. accuracy for particular matrix elements. This potentially enables us to bias the accuracy of bands closer to the Fermi level, reducing the accuracy (and cost) for irrelevant bands. 

The WMCA is highly parallelizable. First of all, each angular momentum channel of the potential, for each atom, is independent and can be calculated in parallel. Once these are calculated and all other common preliminary data are available (e.g. the spherical Bessel transforms of the orbitals), each potential 2CI is also independent and can be parallelized.  However, determining the optimal workload distribution that minimizing communication overhead appears to be complicated. The spherical Bessel transforms are the most expensive part of each 2CI; In future work, I plan to map these onto Hankel transforms that can be calculated with optimized "fast Hankel transform" library routines \cite{secada1999numerical,siegman1977quasi,talman1978numerical}.

Taking position derivatives of the matrix elements, i.e. calculating forces, ought to be straightforward. In spherical coordinates, the radial dependence of matrix elements is contained in the inverse spherical Bessel transforms, eqs. \ref{eq:lambda} and \ref{eq:lambda_pot}. If these are calculated on splines, accurate numerical derivatives are straightforward. Alternatively, simple recursion relations can be used to compute derivatives of the spherical Bessel functions. The angular dependence of the matrix element is contained in the Wigner matrices; if the Wigner matrices are calculated using the algorithm in \ref{appendix:wigner_matrices}, the angular derivatives act on the exponentials and the derivatives can be calculated directly by summing the result. A similar algorithm for derivatives of matrix elements using Wigner matrices has been proposed before \cite{elena2005automatic}.

The WMCA is potentially an $O(n)$ algorithm for matrix element calculations. This is significant because the computational cost of LCAO methods is usually dominated by synthesizing the matrix elements \cite{kohanoff2006electronic,singh2006planewaves}, leading to the active development of other $O(n)$ algorithms \cite{soler2002siesta,lewis2011advances,garcia2020siesta,blum2009ab}. The fact that the WMCA is linear-scaling with respect to the number of atoms follows from the locality of the basis functions: each orbital only overlaps strongly with others located in a relatively small sphere centered on it. Then the multipole expansion of the potential only needs to be calculated within a small sphere centered on each atom and only a fixed, finite set of potential matrix elements needs calculated within each sphere. The range of the orbitals sets the angular momentum cutoff of the potential, $L_\textrm{cut}$, which in turn determines the number of potential 2CI that need calculated; none of these quantities depends on the system size and the WMCA is $O(n)$ in this regard.

I used a dual-space representation of the potential \cite{vandevondele2005quickstep,lippert1997hybrid,kuhne2020cp2k} in the examples above because it was simple and demonstrated the generality of the WMCA. However, the number of plane waves (i.e. grid points) in the potential scales with system size; if a grid representation is used, the WMCA is not $O(n)$. Alternative algorithms are needed for calculating the multipole expansion of the potential without resorting to grids. A special case of an $O(n)$ algorithm for spherical, atom-centered pseudopotentials is presented in the supplementary information \cite{supp}. Alternative linear-scaling algorithms for calculating the multipole expansion and the total energy are the subject of ongoing work. 

The WMCA may also be applicable to a type of tight binding approximation to density functional theory \cite{elstner1998self,elstner2014density,spiegelman2020density,hourahine2020dftb+,hourahine2010dftb+,bannwarth2019gfn2,bannwarth2021extended,grimme2017robust}. These methods are fast because they use the 2CA to approximate the potential matrix elements as 2CI. Parameterized functions are used for the matrix elements and total energy. In the spirit of these methods, it may be possible to devise a minimal pseudoatomic basis with very smooth, short ranged functions, rendering the WMCA very fast. With smooth enough basis functions, the Kohn-Sham total energy can be calculated explicitly while treating the matrix elements as 2CI; this could be very efficient, while retaining transferability. Once the total energy is available, forces and molecular dynamics etc. are possible. The only tunable parameters would belong to the basis set and pseudopotentials, which is also the case with ab-initio pseudopotential methods. Finally, the multipole expansion of the potential encodes most of the physics in the Kohn-Sham Hamiltonian into a relatively small set of local atomic data. Such a representation is amenable to machine learning methods in electronic structure calculations \cite{zhang2022equivariant,li2025enhancing,drautz2019atomic,musil2021physics,dusson2022atomic}. 

\section{Conclusion}

I derived a Wigner matrix based algorithm  (WMCA) for calculating the convolution between arbitrary localized functions. As a special case, I applied the WMCA to tight binding integrals in the 2CA and reproduced the Slater-Koster formulae \cite{slater1954simplified}. I then used an atom-centered multipolar expansion of the full crystal potential to reduce the potential matrix elements into sums over 2CI, each term of which is calculated with the WMCA. I emphasize that the application of the WMCA for band structures is applicable to fully ab-initio calculations in principle. The extension of a tight binding-type 2CI algorithm to full potential calculations using the WMCA will hopefully enable computational methods with efficiency approaching that of advanced total energy tight binding calculations with the transferability approaching that of ab-initio methods. The extension of the WMCA to total energy calculations is the subject of future work.

\section{Code Availability}

A \texttt{python} script to calculate the real basis Wigner-$\D$ matrices (i.e. the rotation coefficients), is provided as supplementary Material. As a byproduct, complex basis Wigner-$D$ matrices and little-$d$ matrices are generated as an intermediate step and can be returned to the user without performing the transformation to the real basis. Thus, the provided script can be used to calculate real or complex basis Wigner matrices for arbitrary $l$. Also provided are \texttt{python} scripts using \texttt{pyvista} to generate the spherical harmonic plots in Figures \ref{fig:fig_1}, \ref{fig:fig_2}, and \ref{fig:high_l}. The \texttt{python} code implementing the WMCA for band structures is still under active development; it is not publicly available yet, but can be retrieved directly from the author on reasonable request.

\section{Acknowledgments}

The author thanks Dmitry Reznik, Beau Sterling, and John Wilson for a critical reading of this manuscript. This work was funded by the U.S. Department of Energy, Office of Basic Energy Sciences, Office of Science, under Contract No. DE-SC0024117.

\appendix

\section{Universal Wigner Matrix Algorithm}\label{appendix:wigner_matrices}

To calculate two-center integrals like eq. \ref{eq:two_center_integral} we want to rotate our real spherical harmonics so that $\vec \Delta = \Delta \vec z$. We need an algorithm that is applicable to arbitrarily large angular momenta. I develop one similar to that of Reference \cite{feng2015high} now.

The real spherical harmonics transform under rotation as
\begin{equation}\begin{gathered}
    X_{lm}(\Omega_{\vec r-\vec \Delta}) = \sum_{m'} \D^{l}_{mm'}(\hat \Delta) X_{lm'}(\Omega_{\vec r - \Delta \vec z})
    \label{eq:real_sph_rotation_2}
\end{gathered}\end{equation} 
with $\Omega_{\vec r-\vec \Delta}$ and $\Omega_{\vec r- \Delta\vec z}$ the angles along the original and new axes respectively specified by vector $\vec \Delta$. $\D^{l}_{mm'}(\hat \Delta)$ is the Wigner-$\D$ matrix element for real spherical harmonics with rotation operator $\R_{\hat \Delta}$. The matrices $\bm \D^{l}(\hat \Delta)$ are real by definition. The real Wigner-$\D$ matrices are related to the usual complex Wigner-$D$ matrices, but the details are left until later. The rotation is specified by Euler angles $(\alpha,\beta,\gamma)$. I use the conventions in Sakurai \cite{sakurai2020modern} for space-fixed $ZYZ$ rotation.

The real spherical harmonics are related to the complex ones by a unitary transformation \cite{blanco1997evaluation},
\begin{equation}\begin{gathered}
    X_{lm}(\Omega_{\vec r}) = \sum_{m'} C^{l}_{mm'} Y_{lm'}(\Omega_{\vec r}) .
    \label{eq:complex_to_real_sph_harm}
\end{gathered}\end{equation}
with $\bm C^{l}$ a unitary matrix that is trivial to populate on a computer \cite{blanco1997evaluation}.

The rotation formula for complex spherical harmonics, analogous to eq. \ref{eq:real_sph_rotation_2}, is
\begin{equation}
\begin{gathered}
    Y_{lm}(\Omega_{\vec r-\vec \Delta}) = \sum_{m'} \bar{D}^{l}_{mm'}(\vec \Delta) Y_{lm'}(\Omega_{\vec r-\Delta \vec z})
\end{gathered}
\end{equation}
where the matrix element is complex conjugated. This comes from the definition \cite{sakurai2020modern} of rotations as
\begin{equation}
\begin{gathered}
    |\Omega_{\vec r-\vec \Delta}\rangle = \R_{\hat \Delta} |\Omega_{\vec r- \Delta \vec z} \rangle 
\end{gathered}
\end{equation}
with $\R_{\hat \Delta}$ the rotation operator. It follows that
\begin{equation}
\begin{gathered}
    \langle Y_{lm} |\Omega_{\vec r-\vec \Delta} \rangle = \sum_{m'} \langle Y_{lm} | \R_{\hat \Delta} |Y_{lm'} \rangle\langle Y_{lm'} | \Omega_{\vec r- \Delta \vec z} \rangle \\
     = \Y_{lm}(\Omega_{\vec r-\vec \Delta}) = \sum_{m'} D^{l}_{mm'}(\hat \Delta) \Y_{lm'}(\Omega_{\vec r- \Delta \vec z}) \\
     \rightarrow Y_{lm}(\Omega_{\vec r-\vec \Delta}) = \sum_{m'} \bar{D}^{l}_{mm'}(\hat \Delta) Y_{lm'}(\Omega_{\vec r- \Delta \vec z}) .
     \label{eq:complex_sph_rotation}
\end{gathered}
\end{equation}
$\bm D^{l}(\hat \Delta)$ is the usual Wigner-$D$ matrix for complex spherical harmonics. Inserting this into the definition of a $\bm \D^{l}(\hat \Delta)$ matrix element,
\begin{equation}\begin{gathered}
    \D^{l}_{mm'}(\hat \Delta) = \langle X_{lm} | \R_{\hat \Delta} | X_{lm'}\rangle  
    = \sum_{MM'} \bar{C}_{mM}^{l} D^{l}_{MM'}(\hat \Delta) C^{l}_{m'M'} 
\end{gathered}\end{equation} 
and 
\begin{equation}\begin{gathered}
    \bm \D^{l}(\hat \Delta) = \bar{\bm C}^{l} \bm D^{l}(\hat \Delta) [\bm C^{l}]^T
\end{gathered}\end{equation} 
 where $[\bm C^{l}]^T$ is the transpose of $\bm C^{l}$ and $\bar{\bm C}^{l}$ its complex conjugate. Then calculating the Wigner-$\D$ matrix is a matter of (i) evaluating the $\bm C^{l}$ matrix \cite{blanco1997evaluation}, (ii) calculating the complex Wigner-$D$ matrix, and (iii) transforming to the real Wigner-$\D$. The hardest part is calculating the Wigner-$D$ matrix, which I do now using a Pauli matrix diagonalization based algorithm similar to that of Reference \cite{feng2015high}.

\subsection{Complex Wigner-$D$ Matrices}

The complex Wigner-$D$ matrix for rotation operator $\R_{\hat \Delta}$ is defined by eq. \ref{eq:complex_sph_rotation} as \cite{sakurai2020modern}
\begin{equation}\begin{gathered}
    D^{l}_{mm'}(\hat \Delta) = \langle Y_{lm} | \R_{\hat \Delta} | Y_{lm'}\rangle 
    = \langle Y_{lm} | e^{-i\alpha \bm L_z}e^{-i\beta \bm L_y}e^{-i\gamma \bm L_z} | Y_{lm'} \rangle 
\end{gathered}\end{equation}
with $(\alpha,\beta,\gamma)$ the Euler angles for the rotation and I used the representation of $\R_{\hat \Delta}$ in terms of angular momentum operators $\bm L_z$ and $\bm L_y$. Acting with the operators
\begin{equation}\begin{gathered}
    D^{l}_{mm'}(\hat \Delta) = e^{-i\alpha m} \langle Y_{lm} | e^{-i\beta \bm L_y}| Y_{lm'} \rangle e^{-i\gamma m'} 
    = e^{-i\alpha m} d^{l}_{mm'}(\beta) e^{-i\gamma m'}  \\
    d^{l}_{mm'}(\beta) = \langle Y_{lm} | e^{-i\beta \bm L_y} | Y_{lm'} \rangle.
\end{gathered}\end{equation}
The matrix $\bm d^{l}(\beta)$ is called the Wigner little-$d$ matrix. $\bm L_y$ is not diagonal, so we can't act it on the states, but we can diagonalize it by similarity transformation $\bm L_y = \bm U^{l} \bm L_z [\bm U^{l}]^\dagger$ and use the identity $e^{\bm U \bm A \bm U^\dagger} = \bm Ue^{\bm A}\bm U^\dagger$ for unitary $\bm U$. $[\bm U^{l}]^\dagger$ is the Hermitian conjugate of $\bm U^{l}$. Then
\begin{equation}\begin{gathered}
    d^{l}_{mm'}(\beta) =  \langle Y_{lm} | \bm U^{l} e^{-i\beta \bm L_z} [\bm U^{l}]^\dagger  | Y_{lm'} \rangle \\
    =  \sum_{M} \langle Y_{lm} | \bm U^{l} |Y_{lM}\rangle \langle Y_{lM} | e^{-i\beta M} |Y_{lM}\rangle \langle Y_{lM} | [\bm U^{l}]^\dagger | Y_{lm'} \rangle \\
    =   \langle Y_{lm} | \bm  U^{l} \bm  K^{l}(\beta)  [\bm  U^{l}]^\dagger | Y_{lm'} \rangle 
\end{gathered}\end{equation}
$ \bm  K^{l}(\beta)$ is a diagonal matrix in the complex spherical harmonic basis. The matrix elements are $e^{-i\beta M}$. E.g. for $l=1$, it is 
\begin{equation}\begin{gathered}
    \bm  K^{l}(\beta) = \begin{pmatrix} e^{-i\beta} & 0 & 0 \\
    0 & 1 & 0 \\
    0 & 0 & e^{i\beta} 
    \end{pmatrix}.
\end{gathered}\end{equation}
Then we can concisely write the $d$ matrix as 
\begin{equation}\begin{gathered}
    \bm  d^{l}(\beta) = \bm U^{l} \bm  K^{l}(\beta) [\bm U^{l}]^\dagger 
\end{gathered}\end{equation}
and, similarly, we can rewrite the $\bm  D^{l}(\hat \Delta)$ matrix as
\begin{equation}\begin{gathered}
    D^{l}_{mm'}(\hat \Delta) =  \langle Y_{lm} | \bm  D^{l}(\hat \Delta) | Y_{lm'} \rangle \\ 
    \bm D^{l}(\hat \Delta) =  \bm  K^{l}(\alpha) \bm U^{l} \bm K^{l}(\beta) [\bm U^{l}]^\dagger \bm  K^{l}(\gamma) ,
    \label{eq:wigner_D_matrix}
\end{gathered}\end{equation}
where I replaced the exponentials by their equivalent diagonal matrices. Explicitly in terms of the other matrix elements, the Wigner-$D$ matrix element is
\begin{equation}\begin{gathered}
     D^{l}_{mm'}(\hat \Delta) = e^{-i\alpha m -i\gamma m'} \sum_M e^{-i\beta M}U^{l}_{mM}  \bar U^{l}_{m'M} .
\end{gathered}\end{equation}
The angular momentum operator is related to the Pauli-$y$ matrix $\bm \sigma^{l}_y$ by $\bm L_y = l \bm \sigma^{l}_y$. Explicit calculation of the Wigner-$D$ matrix requires us to calculate this. We can use the well known formula:
\begin{equation}
\begin{gathered}
    [\bm \sigma^{l}_y ]_{mm'} = \frac{i}{2} (\delta_{m,m'+1} - \delta_{m+1,m'}) \sqrt{(l+1)(m+m'-1)-mm'} .
\end{gathered}
\end{equation}
The matrices $\bm U^{l}$ can be found by diagonalizing $\bm \sigma^{l}_y$.

\begin{figure}
    \centering
    \includegraphics[width=0.5\linewidth]{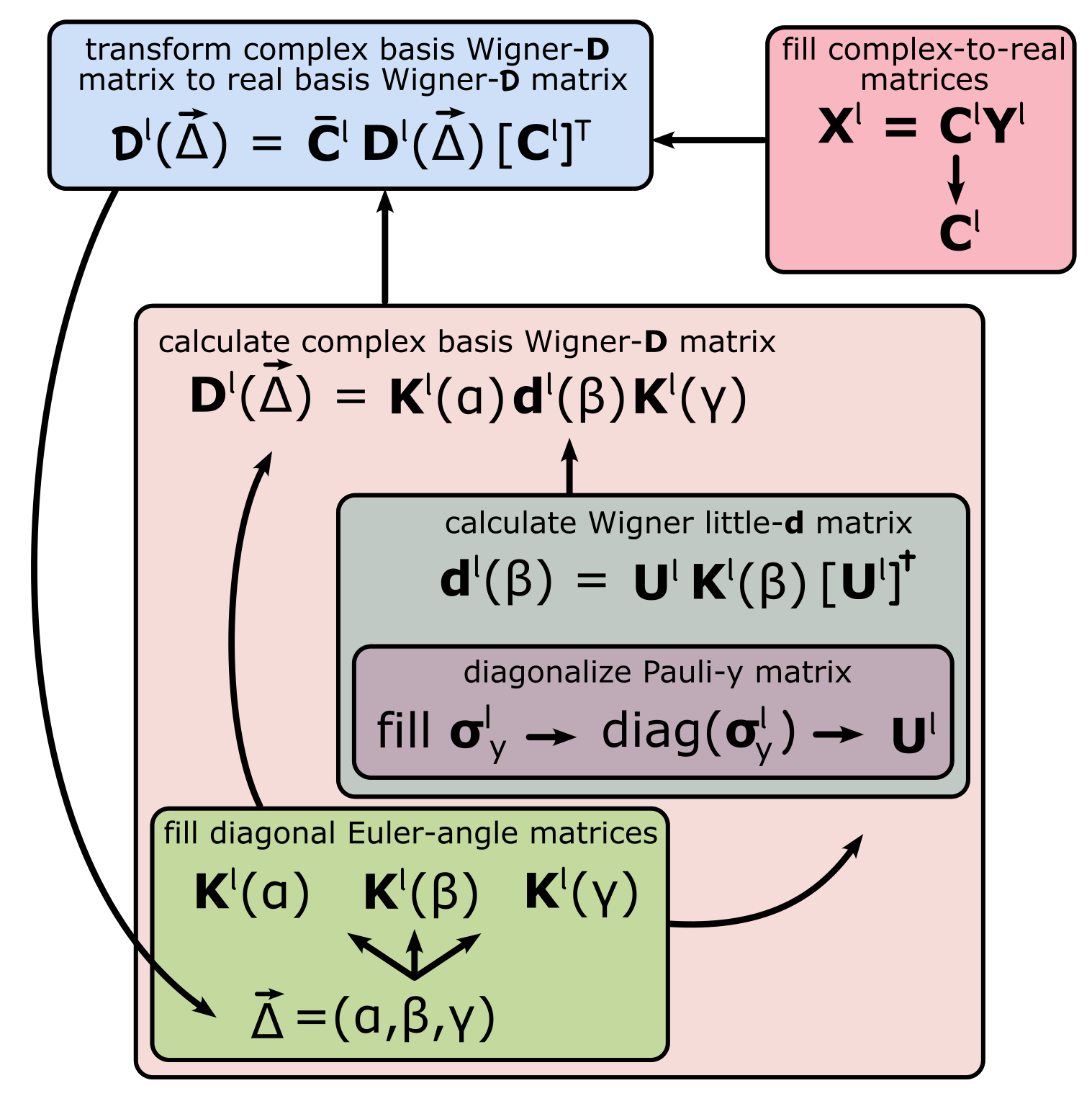}
    \caption{Summary of the algorithm for calculating Wigner matrices for real spherical harmonics. Given a vector $\vec \Delta$, determine the Euler angles $(\alpha,\beta,\gamma)$ and fill the diagonal Euler angle matrices, $\bm K^l(\alpha)$, and so on. Separately, diagonalize the Pauli-$y$ matrix, $\bm \sigma_y^l$. Use the Pauli$-y$ eigenvector matrices, $\bm U^l$, and the $\beta$ Euler angle matrix to calculate the Wigner little-$d$ matrix. Use the $\alpha$ and $\gamma$ Euler angle matrices and the little-$d$ matrix to calculate the Wigner-$D$ matrix. Finally, use the complex to real spherical harmonic transformation matrices $\bm C^l$ to transform the complex basis Wigner-$D$ matrix to the real basis Wigner-$\D$ matrix.}
    \label{fig:algorithm}
\end{figure}

\subsection{The Real Wigner Matrix Algorithm}

Now combine all the steps together (Fig. \ref{fig:algorithm}). In order to calculate two-center integrals, we need to rotate our functions to point along the separation vector $\vec{\Delta}$ which is accomplished as 
\begin{equation}
\begin{gathered}
    H_{lml'm'}(\vec{\Delta}) = \sum_{M} \D^{l}_{mM}(\hat \Delta)\D^{l'}_{m'M}(\hat \Delta) (ll'm) .
\end{gathered}
\end{equation}
See eq. \ref{eq:sk_2CI}. The Wigner-$\D$ matrix elements $\D^{l}_{mm'}(\hat \Delta)$ rotate the spherical harmonics to point along the vector $\vec{\Delta}$. 

To calculate the $\bm  \D^{l}(\hat \Delta)$ matrices, transform from real to complex spherical harmonics: $\bm \D^{l}(\hat \Delta) = \bar{\bm C}^{l} \bm D^{l}(\hat \Delta) [\bm C^{l}]^T$. Calculate $\bm D^{l}(\hat \Delta)$ by diagonalizing the Pauli-$y$ matrix $\bm \sigma^{l}_y$ and transforming the $\bm L_y = \bm U^{l} \bm L_z [\bm U^{l}]^\dagger$ rotation to the $\bm L_z$ basis: $\bm D^{l}(\hat \Delta) = \bm K^{l}(\alpha)\bm U^{l} \bm K^{l}(\beta) [\bm U^{l}]^\dagger \bm K^{l}(\gamma)$ with $\bm K^{l}(\alpha)$ a diagonal matrix of elements $e^{-i\alpha m}$ with $m$ the eigenvalues of $\bm L_z$. Then
\begin{equation}
\begin{gathered}
    \bm \D^{l}(\hat \Delta) = \bar{\bm C}^{l} \bm K^{l}(\alpha) \bm U^{l} \bm K^{l}(\beta) [\bm U^{l}]^\dagger \bm K^{l}(\gamma) [\bm C^{l}]^T
\end{gathered}
\end{equation}
Calculating a $\bm \D^{l}(\hat \Delta)$ matrix requires (i) diagonalizing the Pauli-$y$ matrix, (ii) applying the matrices to $\bm K^{l}(\beta)$, and (iii) transforming to the real spherical harmonic basis. Finally, summing over $M$ requires $2\times \min(l,l')+1$ operations. The required matrix diagonalization and multiplications can be done efficiently using optimized linear algebra libraries.

\begin{figure}
    \centering
    \includegraphics[width=0.5\linewidth]{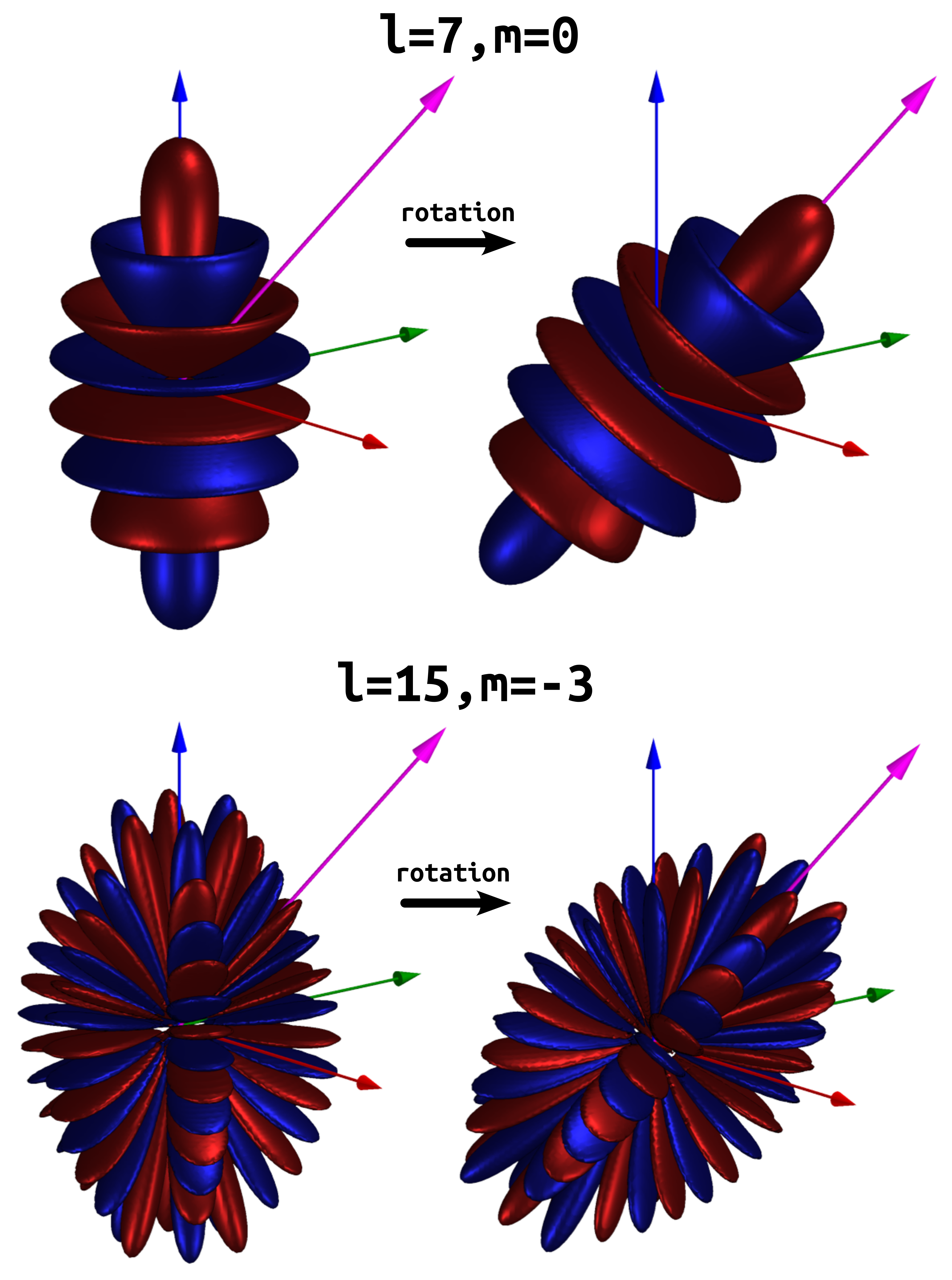}
    \caption{Examples of rotations for high angular momentum (real) spherical harmonics using Wigner-$\D$ matrices. The top row shows rotating an $l=7,~m=0$ real spherical harmonic to point along $\vec{\Delta}=(0.3,0.3,0.5)$. The bottom row shows the same for an $l=15,~m=-3$ function. The rotations were done according to eq. \ref{eq:real_sph_rotation_2}. The plots were made with \texttt{pyvista} \cite{sullivan2019pyvista}.}
    \label{fig:high_l}
\end{figure}

As an explicit test of the applicability to high angular momentum, Wigner-$\D$ matrices are used to rotate high angular momentum functions to point along an arbitrary vector $\vec{\Delta } = (0.3,0.3,0.5)$. Figure \ref{fig:high_l} shows the results. The nature of this algorithm is that any $l$ is possible without implementing new code, which is demonstrated by rotating both the $X_{l=7,m=0}(\Omega)$ and $X_{l=15,m=-3}(\Omega)$ functions (and the $p$ orbitals in Fig. \ref{fig:fig_2}). These were rotated explicitly using eq. \ref{eq:real_sph_rotation_2}. 

\subsection{Analytical Calculation for $l=1$}\label{appendix:analytical}

I now workout the rotation matrices for $s$ and $p$ orbitals analytically. For $s$, the $\bm \D^{0}(\hat \Delta) = 1$ matrix is trivial so I only need to calculate the $\bm \D^{1}(\hat \Delta)$ matrix for $p$ orbitals. In what follows, matrices are written in \emph{ascending} order with respect to $m$. The complex to real spherical harmonic transformation matrix is \cite{blanco1997evaluation} 
\begin{equation}
\begin{gathered}
    \bm C^{l} = \frac{1}{\sqrt{2}} \begin{pmatrix} i & 0 & i \\
    0 & \sqrt{2} & 0 \\
    1 & 0 & -1 
    \end{pmatrix},
\end{gathered}
\end{equation}
the $\bm K^{l}(\alpha)$ matrices are
\begin{equation}
\begin{gathered}
    \bm K^{l}(\alpha) = \begin{pmatrix}
        e^{i\alpha} & 0 & 0 \\
        0 & 1 & 0 \\
        0 & 0 & e^{-i\alpha}
    \end{pmatrix} ,
\end{gathered}
\end{equation}
and the eigenvectors of the $\bm L_y$ operator are
\begin{equation}
\begin{gathered}
    \bm U^{l} = \frac{1}{\sqrt{2}} \begin{pmatrix} \frac{1}{\sqrt{2}} & 1 & -\frac{1}{\sqrt{2}} \\
    i & 0 & i \\
    -\frac{1}{\sqrt{2}} & 1 & \frac{1}{\sqrt{2}} 
    \end{pmatrix} .
\end{gathered}
\end{equation}
Carrying out the matrix multiplications (I use shorthand $\cos\alpha=c_\alpha$, $\sin\alpha=s_\alpha$, and so on)
\begin{equation}
\begin{gathered}
    \bm \D^{1}(\hat \Delta) = 
    \begin{pmatrix} c_\alpha c_\gamma - s_\alpha c_\beta s_\gamma & s_\alpha s_\beta & c_\alpha s_\gamma + s_\alpha c_\beta c_\gamma \\
    s_\beta s_\gamma & c_\beta & - s_\beta c_\gamma \\
    - s_\alpha c_\gamma -c_\alpha c_\beta s_\gamma   & c_\alpha s_\beta & - s_\alpha s_\gamma + c_\alpha c_\beta c_\gamma
    \end{pmatrix} .
\end{gathered}
\end{equation}
Permuting axes to align with the conventional $x,y,z$ order (in the real spherical harmonic basis, this is ordered according to $m = (-1,0,1) = (y,z,x)$), this agrees with the known Euler rotation matrix for $ZYZ$ rotation in three dimensions.


\end{document}